\documentclass[useAMS,usenatbib]{mn2e}

\usepackage{graphicx}
\usepackage{subfigure}
\usepackage{amssymb}
\usepackage{amsmath}
\usepackage{tabularx}
\usepackage{float}
\usepackage{placeins}
\usepackage{adjustbox}

\def\aap {{A\&A}}

\def\apjl {{ApJL}}

\def\apjs {{ApJS}}

\title[UV Escape Fractions from GMCs]{Simulating the UV Escape Fractions from Molecular Cloud Populations in Star-forming Dwarf and Spiral Galaxies}

\author[C.S.\ Howard, R.E.\ Pudritz, W.E.\ Harris, \& R.S. Klessen]{Corey \ S. \ Howard$^{1}$\thanks{E-mail: howardcs@mcmaster.ca}, Ralph \ E.\ Pudritz$^{1,2}$, William \ E.\ Harris$^{1}$, Ralf \ S. \ Klessen$^{3,4}$ \\
$^{1}$Department of Physics and Astronomy, McMaster University, 1280 Main St.~W, Hamilton, ON L8S 4M1, Canada\\
$^{2}$Origins Institute, McMaster University, 1280 Main St.~W, Hamilton, ON L8S 4M1, Canada\\
$^{3}$ Universit\"{a}t Heidelberg, Zentrum f\"{u}r Astronomie, Institut f\"{u}r Theoretische Astrophysik, Albert-Ueberle-Stra{\ss}e 2, 69120 Heidelberg, Germany\\
$^{4}$ Universit\"{a}t Heidelberg, Interdisziplin\"{a}res Zentrum f\"{a}r Wissenschaftliches Rechnen, Im Neuenheimer Feld 205, 69120 Heidelberg, Germany
}

\begin{document}
\bibliographystyle{mn2e}

\date{10 October 2017}

\pagerange{\pageref{firstpage}--\pageref{lastpage}} \pubyear{2017}

\maketitle

\label{firstpage}

\begin{abstract}

\noindent\noindent The escape of ultraviolet photons from the densest regions of the interstellar medium (ISM) --- Giant Molecular Clouds (GMCs) --- is a poorly constrained parameter which is vital to 
understanding the ionization of the ISM and the intergalactic medium. We characterize the escape fraction, f$_{\text{esc,GMC}}$, from a suite of individual GMC simulations with masses in the range 10$^{4-6}$ 
M$_{\odot}$ using the adaptive-mesh refinement code FLASH. We find significantly 
different f$_{\text{esc,GMC}}$ depending on the GMC mass which can reach $>$90\% in the evolution of 5$\times$10$^4$ and 10$^{5}$ M$_{\odot}$ clouds or remain low at $\sim$5\% for most of the lifetime of more massive GMCs. All 
clouds show fluctuations over short, sub-Myr timescales produced by flickering HII regions. We combine our results to calculate the total escape fraction (f$_{\text{esc,tot}}$) from GMC populations 
in dwarf starburst and spiral galaxies by randomly drawing clouds from a GMC mass distribution (dN/dM$\propto$M$^{\alpha}$, where $\alpha$ is either -1.5 or -2.5) over 
fixed time intervals. We find typical f$_{\text{esc,tot}}$ values of 8\% 
for both the dwarf and spiral models. The fluctuations of f$_{\text{esc,tot}}$, however, are much larger for the dwarf models with values as high as 90\%. The photons escaping from the 
5$\times$10$^4$ and 10$^{5}$ M$_{\odot}$ GMCs are the dominant contributors to f$_{\text{esc,tot}}$ in all cases. We also show that the accompanying star formation rates (SFRs) of our model 
($\sim$2$\times$10$^{-2}$ and 0.73 M$_{\odot}$yr$^{-1}$) are consistent with observations of SFRs in dwarf starburst and spiral galaxies, respectively. 

\end{abstract}


\section{Introduction} 

The emission, absorption, and reprocessing of ultraviolet (UV) photons produced by massive stars are important processes both within a galaxy and in the intergalactic medium (IGM). 
On galactic scales, these photons contribute to the interstellar radiation field (ISRF), first characterized by \citet{Habing}. This field is responsible for the thermal, chemical, and ionization state 
of the interstellar medium (ISM) \citep{Draine2011}. Despite their short lifetimes, massive stars in the range of 10-100 M$_{\odot}$ are a strong contributor to
the ISRF due to their high UV luminosities. Moreover, it has been shown that the isolated field O-star population of the Milky Way (MW) is not sufficient to maintain the ionization of the Diffuse 
Ionized Gas (DIG) layers above and below the galactic plane \citep{Reynolds3}. Instead, a significant portion must be contributed via young O-stars still located in their birth cluster.
The HII regions surrounding these massive stars must, therefore, be leaking photons into the ISM. How, and the degree to which, UV photons propagate from massive stars out of their gaseous birth sites is therefore crucial 
for the large-scale structure of the ISM.

Once a photon escapes a galaxy, it can interact with the IGM and, at high redshifts (z$\geq$6), contribute to cosmic reionization \citep{Robertson2010}. However, the exact fraction of UV photons that escape 
from a galaxy, f$_{\text{esc}}$, is poorly constrained and is likely a function of the host galaxy's properties. For example, estimates of f$_{\text{esc}}$ for high redshift galaxies measured 
via the Lyman continuum range from 7\% \citep{Siana2015} to 30\% \citep{Nestor2013}. Recent estimates of f$_{\text{esc}}$ calculated via HII region mapping for the Small Magellanic Cloud (SMC) and the Large Magellanic 
Cloud (LMC) suggest values of 4\% and 11\%, respectively \citep{Pelle2012}.

It has been shown that galaxies which host active galactic nuclei (AGN) 
are not sufficient to drive cosmic reionization \citep{Fan2006,Robertson2013}. Instead, dwarf galaxies are thought to be important contributors. Estimates suggest that $\sim$40\% of the 
total ionizing photons required for cosmic reionization may be produced in dwarfs \citep{Wise2014}.

Since the masses of these dwarfs are small --- as low as $\sim$10$^8$ M$_{\odot}$ --- they are not observable at high redshifts. More importantly, the IGM opacity makes 
direct observations of LyC radiation from galaxies at z$>$4 difficult and at z$>$6, impossible \citep{Inoue,Rutkowski}. Simulations are therefore required to 
constrain f$_{\text{esc}}$. As with the observations mentioned above, there is a large variation between quoted values obtained from simulations. For example, the predicted f$_{\text{esc}}$ for high redshift 
dwarfs ranges from $<$1\% \citep{Pard2011} to $>$10\% \citep{Raz2010,Ferrara2013,Pard2015,Schauer}. Moreover, simulations indicate that f$_{\text{esc}}$ can range over several orders of magnitude throughout an individual galaxy's 
evolutionary history \citep{Pard2011}.
 
The distribution of dense gas in a galaxy, and the treatment of star formation within that gas, is one of the main constraints on modeling f$_{\text{esc}}$ \citep{Pard2011}. Thus detailed studies aimed at 
modeling f$_{\text{esc}}$ from Giant Molecular Clouds (GMCs) --- the densest regions of the ISM and the hosts to massive star formation --- can provide 
important constraints on the global f$_{\text{esc}}$ from a galaxy. This problem is complicated, however, by the complex internal structure of GMCs, the variation in physical 
conditions from cloud to cloud, and the addition of several physical processes at the onset of star formation.

GMCs as a whole consist of dense filaments formed as a product of supersonic turbulence \citep{BertoldiMckee, Lada2003, MaclowKlessen, McKee2007, PPVI, Klessen2016}. 
The mass of individual clouds, however, can vary over several orders of magnitude. Within the MW, GMCs range from 10$^{4-7}$ M$_{\odot}$ \citep{Fukui} with a 
powerlaw mass distribution of $dN/dM \propto M^{-1.5}$ \citep{Solomon1987, Roso2005}. The virial parameter of GMCs (i.e.. the ratio of internal kinetic energy to gravitational 
potential energy) also covers a wide range, from significantly bound to unbound \cite[$\alpha$ = 0.5 - 5, ][]{Blitz}. As shown in \citet{Howard2016}, the virial parameter 
plays a central role in determining the star formation efficiency (SFE) of a GMC. 

At the onset of star cluster formation --- typically occurring at the intersection of filaments for massive clusters \citep{Balsara,Schneider2012,Kirk2013} --- energy and momentum are 
imparted to the gas by a variety of feedback processes from newly formed stars. Stellar winds \citep{Dale2008,Gatto2017}, protostellar jets \citep{Li2006,Maury2009,Federrath2014}, radiative 
feedback \citep{Dale2005,Peters2010,Klassen2012,Howard2016}, and supernovae feedback \citep{Rogers,Fierlinger} are several examples of these processes. Radiative feedback is 
particularly important for clusters hosting massive star formation \citep{Murray2010,Bate2012,Rahner}. Ultraviolet photons produced by massive stars heat and ionize 
the surrounding gas resulting in the formation of HII regions. The interaction of photons with dust grains can also impart momentum into the gas and drive outflows.  

In \citet{Howard2017-2}, we examined the role that radiative feedback plays in controlling the SFE and star formation rates (SFR) of young, filamentary GMCs over a mass range of 10$^{4-6}$ M$_{\odot}$. We completed a suite of 5 simulations of turbulent GMCs using the code FLASH which combined sink particles to represent star-forming clusters and 
a raytracing scheme to complete the radiative transfer. We found that the inclusion of radiative feedback lowered the SFEs for all clouds, but GMCs in the range of 5$\times$10$^4$ to
10$^5$ M$_{\odot}$ were affected the most. In that mass range, the energy injected by radiative feedback outweighs the gravitational potential energy of the cloud, resulting in 
nearly complete ionization by $\sim$5 Myr.

In this paper, we take the next step and compute the UV escape fraction from the same suite of simulations in order to put further constraints on the global escape fraction from the observed mass spectrum of clouds in entire galaxies. In Section 
\ref{sec:method2}, we provide an overview of our numerical methods and the details of the escape fraction calculation.

We first discuss (Section \ref{Results51}) the escape fractions from individual GMCs (f$_{\text{esc,GMC}}$). We find that cloud mass plays an important role in determining f$_{\text{esc,GMC}}$ in this mass range. For clouds of mass 
5$\times$10$^4$ and 10$^5$ M$_{\odot}$, the final f$_{\text{esc,GMC}}$ are 90\%, and 100\% at 5 Myr. Still more massive clouds have correspondingly low f$_{\text{esc,GMC}}$ which do not 
exceed $\sim$12\%. Regardless of GMC mass, there are large fluctuations in f$_{\text{esc,GMC}}$ over short timescales due to dynamic HII regions which grow and 
shrink rapidly depending on the local conditions surrounding the luminous clusters embedded within the GMCs.

We then present a model (Section \ref{Results52}) for combining our results for individual GMCs in order to represent the escape fraction from a \textit{population} of GMCs (f$_{\text{esc,tot}}$). This model involves 
drawing clouds at random from a GMC mass distribution over fixed time intervals at random, evolving the clouds over an assumed lifetime, and calculating f$_{\text{esc,tot}}$ from the entire population. We 
present two realizations of this model --- one representative of a dwarf starburst galaxy, and one for a normal spiral-type galaxy. We note that these calculations are meant to represent 
the escape fraction from GMCs and not an entire galaxy. To treat the latter problem, a detailed model of the galactic ISM as a whole is required. Moreover, a substantial fraction ($\sim$30\%) of the 
UV photons escaping the galactic disk are absorbed by the gas in the halo of the galaxy \citep{Prochaska,Trebitsch}. Quantifying f$_{\text{esc,GMC}}$, however, is the 
important first step towards understanding galactic escape fractions.

We find typical f$_{\text{esc,tot}}$ of $\sim$8\% for both the dwarf and spiral models. However, there are strong variations in f$_{\text{esc,tot}}$ with time, particularly 
for the dwarf models which cover the range of 0-90\%. We also present the resulting SFRs and find typical values of $\sim$2$\times$10$^{-2}$ and 0.73 M$_{\odot}$yr$^{-1}$
for the dwarf and spiral model which are consistent with observed values \citep{Gao2004,McQuinn,McQuinn2,MilkyWay}. We discuss the consequences of these results for galactic evolution. 

\section{Numerical Methods} \label{sec:method2}

\subsection{Simulation Details}

We briefly describe the details of our numerical methods and simulation suite below. For more detail, we refer the reader to \citet{Howard2017-2}.

We have completed a suite of GMC models having masses of 10$^4$, 5$\times$10$^{4}$, 10$^5$, 5$\times$10$^{6}$, and 10$^6$ M$_{\odot}$ using the Adaptive Mesh Refinement (AMR)
code FLASH \citep{Fryxwell2000}. FLASH includes modules to treat self-gravity, cooling via molecular lines and dust \citep{Banerjee2006}, radiative transfer, and star formation --- all
of which are included in our simulations.

To isolate the effects of varying GMC mass, we fixed the initial average density to 100 cm$^{-3}$ and the initial virial parameter ($\alpha = 2E_{kin}/E_{grav}$, where $E_{kin}$ is 
the internal kinetic energy and $E_{grav}$ is the gravitational potential energy) to 3 for all models. The density
profile is uniform in the inner half of the cloud and decreases as r$^{-3/2}$ in the outer half. Following the method of \citet{Girichidis2011}, we overlay the clouds with
an initial Burgers turbulent velocity spectrum that contains a natural (1:3) mixture of compressive to solenoidal modes. The turbulence is not driven and decays with time. The initial Mach number of the turbulence differs depending on the cloud mass
in order to keep the virial parameter constant across the models. The resolution also depends on the GMC mass and ranges from 0.13 to 0.31 pc (see Table 1 of \citet{Howard2017-2} for a full list of our model parameters).

A hybrid-characteristics raytracing method developed by \citet{Rijkhorst} and adapted for star formation simulations by \citet{Peters2010} is used to treat radiative transfer. Both
ionizing and non-ionizing radiation are included in this scheme. The ionization equations are solved via the DORIC package \citep{Frank1994} assuming hydrogen is the only gas 
component. The Planck mean opacities taken from \citet{Pollack1994} are used for non-ionizing radiation. We include radiation pressure by adopting a single UV opacity of 
$\kappa$ = 775 cm$^2$ g$^{-1}$ \citep{LiDraine2001} that is scaled by the neutral fraction of the gas such that fully ionized regions have zero opacity. The corresponding 
radiative force per unit mass is given by,

\begin{equation}
F = \kappa \frac{L}{c}\frac{e^{-\tau_{uv}}}{4\pi r^2}
\end{equation}

\noindent where $\kappa$ is the opacity to UV radiation, $c$ is the speed of light, $L$ is the source luminosity, $r$ is the distance between the source and the cell, and $\tau_{uv}$ is the optical depth between the source 
and the cell calculated using the raytracer. Radiation pressure is thought to be the dominant form of feedback for GMCs hosting massive cluster formation \citep{Murray2010} which
has been shown to be the case in 10$^7$ M$_{\odot}$ GMCs (Howard, Harris, \& Pudritz, in prep.).

The radiative transfer scheme is coupled to the sink particle method implemented by \citet{Federrath2010}. We use these sink particles to represent star-forming clusters. 
A density threshold for cluster formation of 10$^4$ cm$^{-3}$ is adopted in order to be consistent with observations of star-forming clumps \citep{Lada2003}.

We have developed a subgrid model for how star formation proceeds in these clusters 
\cite[for full details and tests of our model, see][]{Howard2014}. We divide each cluster's
mass into two components; the mass contained in fully formed stars, and the remaining gas mass. Every 0.36 Myr, corresponding to the free fall time
of the gas at our adopted threshold density, 20\% of the remaining gas mass is converted to stars by randomly sampling a \citet{Chabrier2005} IMF. Any gas accreted by the 
cluster from the surrounding GMC is available for future star formation.

The stars formed in each cluster are recorded and the total luminosity of a cluster is the sum of its stellar components. We use analytic fit formulae from \citet{Tout1996} to determine 
each star's total luminosity and ionizing luminosity. These fits were developed for main sequence stars so we therefore neglect the effects of protostellar evolution. The 
radiative properties of each cluster are passed to the raytracer to complete the radiative transfer.

We note that the radiative transfer scheme only tracks the total flux of ionizing photons in each cell and not the net direction of the flux. This leads to some approximations
when calculating the f$_{\text{esc,GMC}}$ --- the escape fraction from a single GMC --- from the cloud, as discussed below. 

\subsection{Escape Fraction Calculation}

To calculate f$_{\text{esc,GMC}}$, we first extract a spherical surface with a radius corresponding to the initial radius of the cloud. We use the marching cube 
algorithm implemented in the YT analysis toolkit \citep{YT} to perform the surface extraction. This algorithm identifies isocontours of the provided quantity (in our 
case the distance from the centre of the simulation volume) and represents these contours as a collection of triangles which, together, represent the extracted surface. There 
can be multiple triangles within an individual grid cell depending on the geometry and the requested isocontour surface.

The total number of photons crossing the extracted surface is then calculated. The three dimensional positions of all surface triangles are known as well as the flux of photons
through each element. In order to calculate the outward flux of photons, the direction of the flux is required. As mentioned above, the radiative transfer scheme only tracks
the magnitude of the flux in each grid cell. We therefore make the assumption that all photons are generated at the centre of luminosity (i.e. the luminosity weighted average 
cluster position). The direction of the flux across each surface element is then known, and the total number of photons crossing the surface is given by

\begin{equation}
N = \sum\limits_{i} \vec{F_{i}}\cdot d\vec{A_i}
\end{equation}

\noindent \noindent where $\vec{F_i}$ is the vector flux of ionizing photons across surface element $i$, and $d\vec{A_i}$ is its corresponding area vector. The escape fraction of UV photons is then 
calculated by dividing $N$ by the total number of UV photons being produced by all clusters.

\begin{figure*}
\centering
\begin{tabular}{ c c }
  \includegraphics[width=0.45\textwidth]{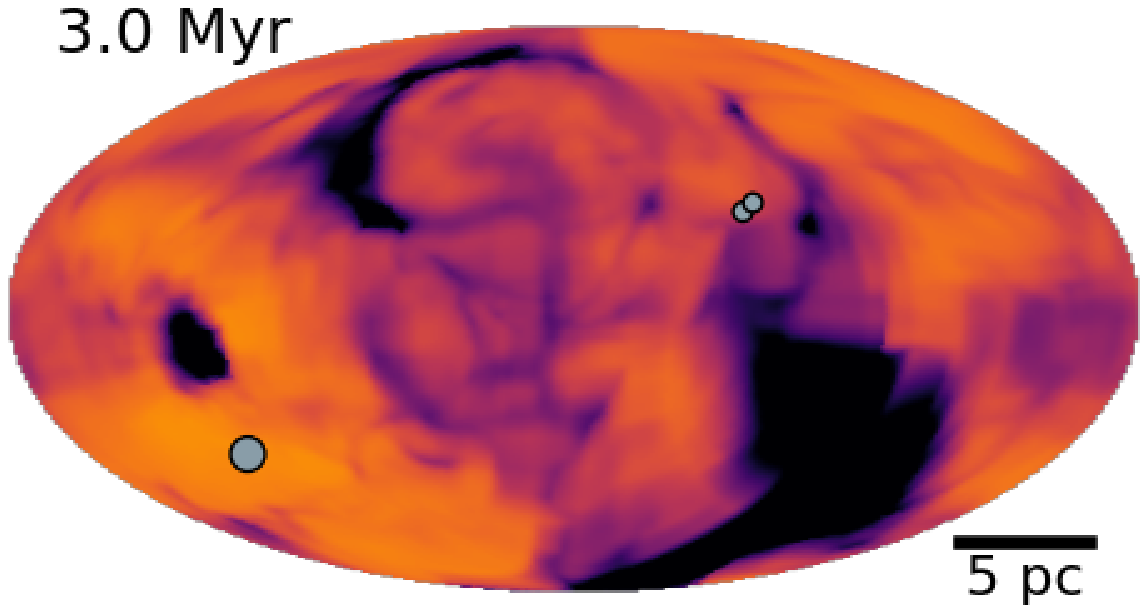} & \includegraphics[width=0.45\textwidth]{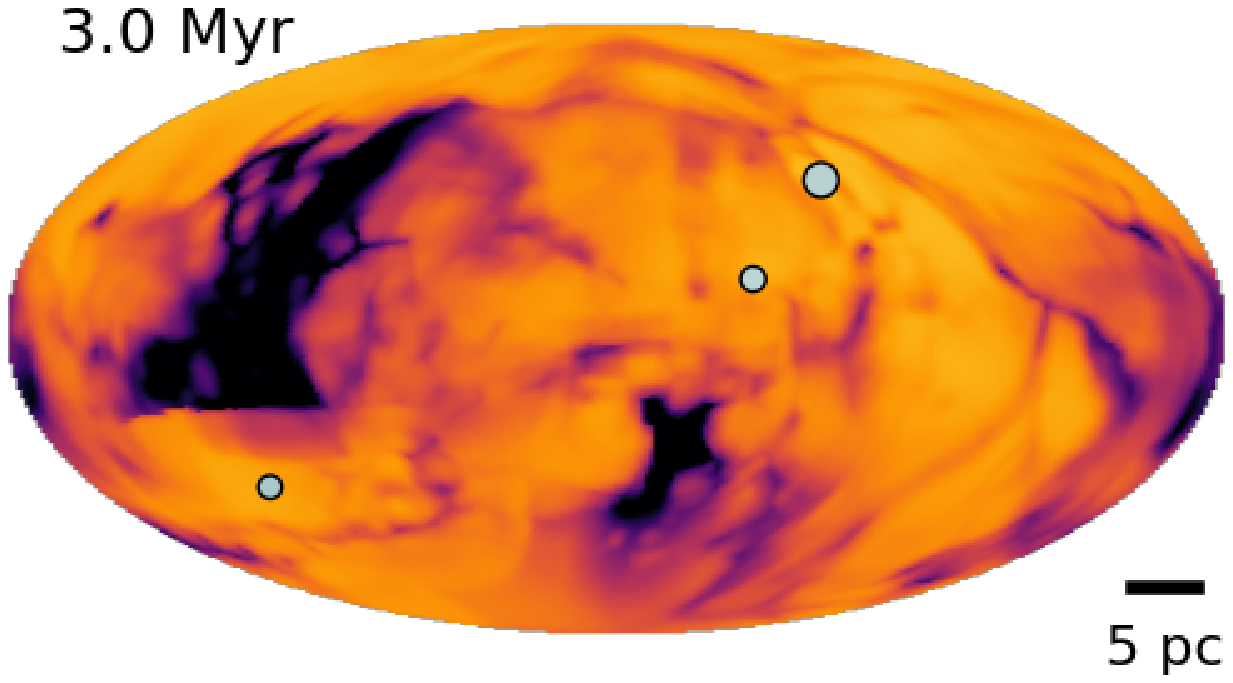} \\
  \includegraphics[width=0.45\textwidth]{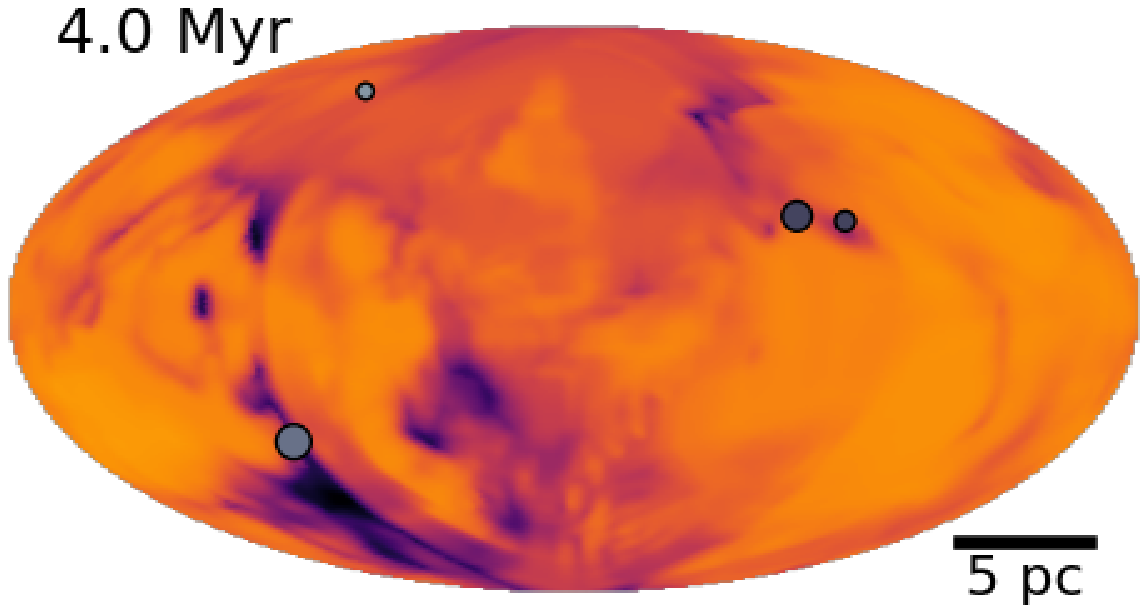} & \includegraphics[width=0.45\textwidth]{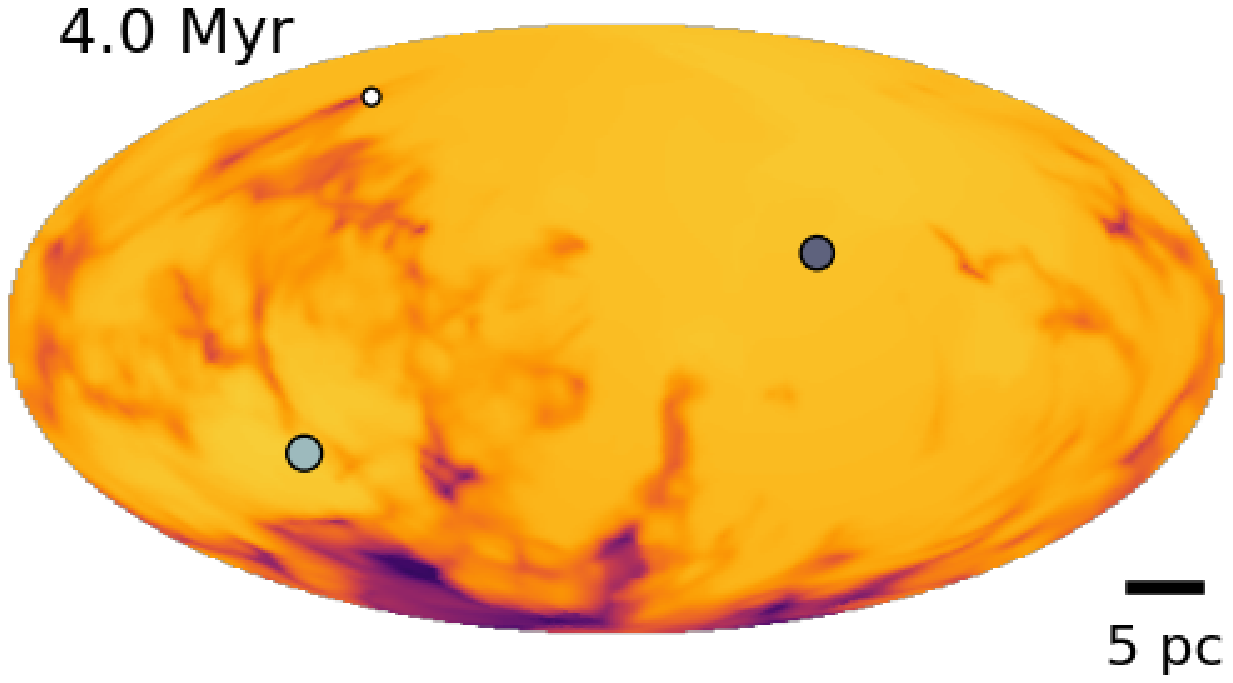} \\
  \includegraphics[width=0.45\textwidth]{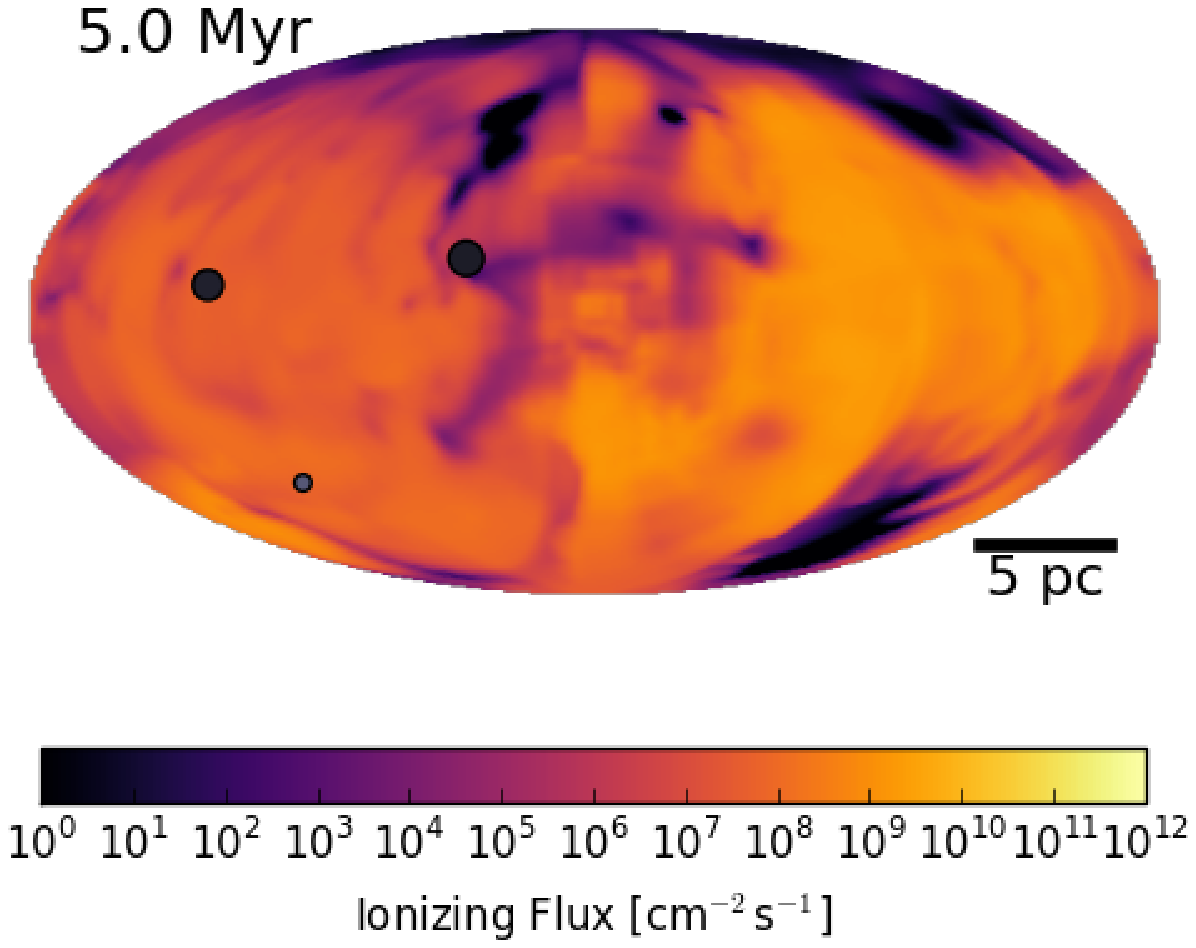} & \includegraphics[width=0.45\textwidth]{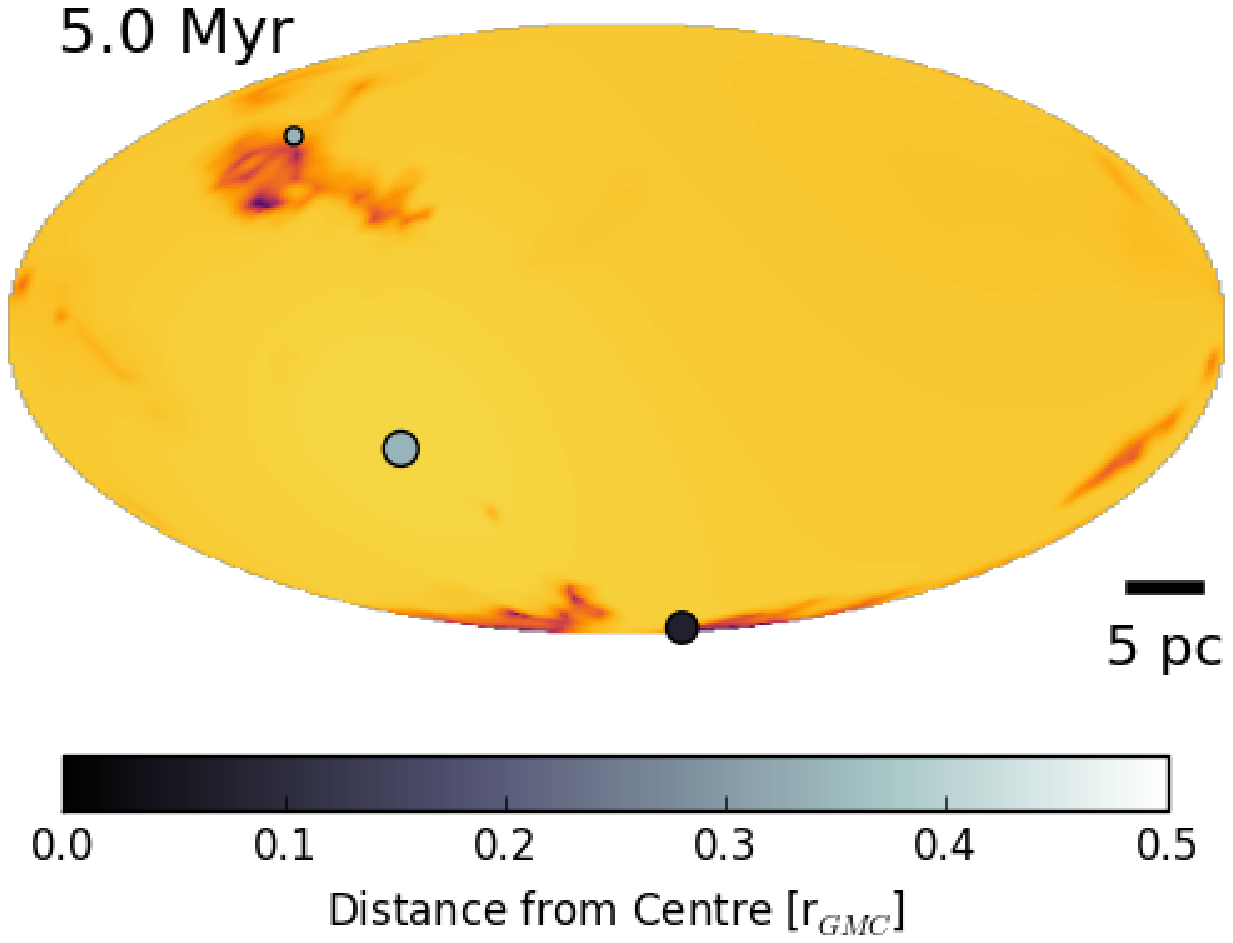} \\
\end{tabular} 
\caption{The flux of ionizing photons across the extracted spherical surface for the 10$^4$ (left) and 10$^5$ (right) M$_{\odot}$ GMCs. Circles represent 
the luminous clusters which are plotted at their closest location to the sphere, scaled in size by their ionizing luminosity, and coloured based on their 
distance (in units of GMC radii) from the centre of the cloud. For reference, the GMC radii for the 10$^4$ and 10$^5$ M$_{\odot}$ GMCs are 7.7 and 16.5 pc. The 
corresponding f$_{\text{esc,GMC}}$ values for the 10$^4$ M$_{\odot}$ GMC, in order of increasing time, are 6.1\%, 11.4\%, and 7.1\%. For the 10$^5$ M$_{\odot}$ GMC, 
the values are 12.2\%, 40.3\%, and 92.6\%.}
\end{figure*}

We note that this method has been improved from what was presented in \citet{Howard2017}. In that work, we did not consider a direction for the flux and simply took $d\vec{A_i}$
to be the area of each grid cell viewed face-on. As we will show below, this led to an overestimate of f$_{\text{esc,GMC}}$, and when applied to our new suite of simulations 
resulted in nonphysical escape fractions that were greater than 1. The improved method produces more accurate estimates, and we quantify the errors 
introduced by our assumption that all radiation is generated at the centre of luminosity in the next Section. 

\section{Individual GMC Escape Fractions} \label{Results51}

Before discussing f$_{\text{esc,GMC}}$ from clouds in a range of masses, we produce visualizations of the ionizing photon flux 
across our extracted surfaces using Hammer projection maps, described in \citet{Howard2017}, in Figure 5.1. Here, we only
produce maps for the 10$^4$ and 10$^5$ M$_{\odot}$ clouds and refer the reader to \citet{Howard2017} for the 10$^6$ M$_{\odot}$ model.

\begin{figure*}
\centering
\includegraphics[width=0.98\textwidth]{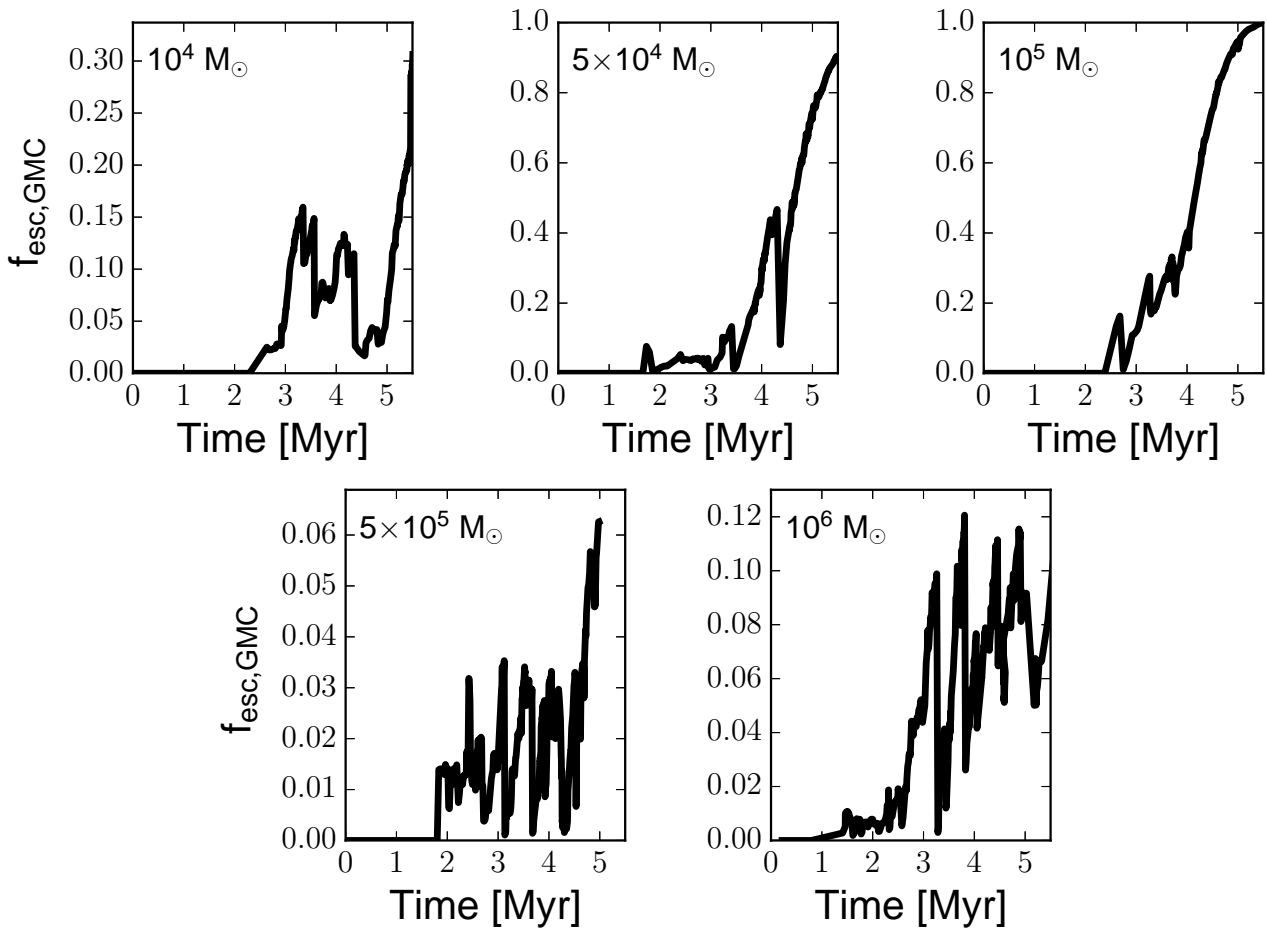}
\caption{The escape fraction from our 5 GMC models of varying mass. The highest f$_{\text{esc,GMC}}$ values are seen in the 5$\times$10$^4$ and 10$^5$ M$_{\odot}$ clouds, both of which are nearly
fully ionized at the end of the simulation. Note that the vertical scales differ between plots.}
\label{fig:grid}
\end{figure*}

A Hammer projection, an equal-area projection which reduces 
the distortion in the outer regions of the map, was used to project the flux of photons across the extracted surface. We also include the positions of the luminous clusters as circles, scaled in size by their ionizing luminosity, which are 
plotted at their closest position to the spherical surface. The clusters are coloured by their distance from the centre of the GMC in units of GMC radii. We note that there are more clusters present in the simulation but small clusters that do not produce 
significant UV flux are not included in the radiative transfer calculation to reduce the overall computational time. The physical sizes of the maps are also different between the 
10$^4$ and 10$^5$ M$_{\odot}$ GMCs so a 5 pc scale bar is included for reference.

Starting at 3 Myr, both GMC models show regions of high and low flux ranging over $\sim$8 orders of magnitude. This is due to the large amount of neutral gas 
still present in the simulation which absorbs photons as they propagate outwards to the surface. The 
clumpy and filamentary nature of the intervening neutral gas is responsible for the appearance of the dark regions.

Significant differences between the two simulations are clearly evident at 5 Myr. While the 10$^4$ M$_{\odot}$ simulation still has regions of low flux, the 10$^5$ M$_{\odot}$
GMC has a nearly uniform flux of $\sim$10$^{11}$ cm$^{-2}$s$^{-1}$. As discussed in \citet{Howard2017-2}, this is due to the cloud being nearly fully ionized
at 5 Myr, meaning the opacity to ionizing luminosity is low and nearly all photons are crossing the plotted surface. The same is true for the 5$\times$10$^4$ M$_{\odot}$ GMCs
which is not plotted here. 

\begin{table*}
\centering
\begin{tabular}{|c|c|c|c|c|}
\hline
\textbf{GMC Mass [M$_{\odot}$]} & \textbf{Average f$_{\text{esc,GMC}}$ [\%]} & \textbf{Net f$_{\text{esc,GMC}}$ [\%]} & \textbf{Peak f$_{\text{esc,GMC}}$ [\%]} & \textbf{Final f$_{\text{esc,GMC}}$ [\%]} \\ \hline
10$^4$ & 9.9 & 12 & 31 & 31 \\ 
5$\times$10$^4$ & 46 & 65 & 90 & 90 \\ 
10$^5$ & 50 & 63 & 100 & 100 \\ 
5$\times$10$^5$ & 1.8 & 2.3 & 6.3 & 6.3 \\ 
10$^6$ & 5.6 & 8.1 & 12 & 9.2 \\ \hline
\end{tabular}
\caption{Escape fraction results for individual GMCs of different masses.}
\end{table*}
On the other hand, the two highest mass clouds (5$\times$10$^5$ and 10$^6$ M$_{\odot}$) are characterized by large variations throughout their entire evolution and show 
only a modest increase in f$_{\text{esc,GMC}}$. The 10$^6$ M$_{\odot}$ model shows two particularly large decreases in f$_{\text{esc,GMC}}$ at 3 and 3.75 Myr, the first of which drops 
from 10\% to $<$1\%. Over the 5 Myr of evolution, f$_{\text{esc,GMC}}$ only rises to a maximum of 12\%. The 5$\times$10$^5$ M$_{\odot}$ oscillates around an f$_{\text{esc,GMC}}$ of 2\% from 2 Myr 
to 4.5 Myr and only reaches a maximum of 6\% in the last 0.5 Myr of evolution. 

The emergence of a fully ionized GMC by $\sim$5 Myr is unique to the 5$\times$10$^4$ and the 10$^5$ M$_{\odot}$ model. For the lowest mass GMCs, there 
is not a large enough population of massive stars to appreciably affect the cloud. On the other hand, the largest clouds remain mostly neutral and bound throughout 
their evolution. The intermediate mass clouds produce enough massive stars to globally unbind the cloud via radiative feedback alone which results in a fully ionized cloud at 
late times.

We plot f$_{\text{esc,GMC}}$ for our five simulated GMCs in Figure \ref{fig:grid}. We adopt the subscript GMC to distinguish between the escape fractions from a population of clouds discussed in 
the next Section. For the reasons discussed above, f$_{\text{esc,GMC}}$ is highest for the 5$\times$10$^4$ and 10$^5$ M$_{\odot}$ 
GMCs. We can quantify our errors based on the 10$^5$ M$_{\odot}$ cloud which, at late times, has f$_{\text{esc,GMC}}$ of 1.07 rather than unity. We therefore have an absolute measurement error in f$_{\text{esc,GMC}}$ of 
$\pm$0.07. As discussed in Section \ref{sec:method2}, the largest source of error is the assumption that all radiation originates at the centre of luminosity.

The final (and average) f$_{\text{esc,GMC}}$, in order of ascending GMC mass, are 31\% (10\%), 90\% (38\%), 100\% (49\%), 6\% (2\%), and 9\% (6\%). The net and peak f$_{\text{esc,GMC}}$ values are also listed in Table 1.
We calculate the average f$_{\text{esc,GMC}}$ by averaging the instantaneous points presented in Figure \ref{fig:grid}. This gives a rough estimate of what the typical observed
f$_{\text{esc,GMC}}$ would be for a given GMC mass. We note that the average f$_{\text{esc,GMC}}$ for the 10$^6$ M$_{\odot}$ GMC has dropped from 15\% in \citet{Howard2017} due to the newly implemented method.
The net f$_{\text{esc,GMC}}$, defined as the total number of photons that escape the GMC over its lifetime divided by the total number of photons generated, is slightly 
higher than the average f$_{\text{esc,GMC}}$.

All models show a high degree of variability. The origin of this variability, discussed in \citet{Howard2017}, is directly related to variable HII region sizes. As an HII region 
grows, the total column density of neutral material between the clusters and the extracted surface decreases resulting in a higher fraction of photons reaching the surface. Due 
to the turbulent nature of that gas combined with the dynamical motions of the clusters, the conditions locally surrounding the clusters can vary. When a cluster enters a region 
of high density, the HII region can be shielded from further radiation which leads to its collapse as the gas recombines. This leads to the sharp declines seen in f$_{\text{esc,GMC}}$.
Fluctuating HII regions have been seen in both observations \citep{DePree2014,DePree2015} and other simulations \citep{Peters2010,UCHII,Madrid2011,Klassen2012}.

Generally, the behaviour of f$_{\text{esc,GMC}}$ follows one of two trends depending on GMC mass --- a smoothly increasing f$_{\text{esc,GMC}}$ at late times with large variations superimposed at early times, 
or a highly variable f$_{\text{esc,GMC}}$ which oscillates around a mean value. A clear example of the first trend is the 5$\times$10$^4$ M$_{\odot}$ cloud. After approximately 3 Myr, f$_{\text{esc,GMC}}$ 
begins to increase eventually reaching 90\% at the end of the simulation. At $\sim$4 Myr, f$_{\text{esc,GMC}}$ drops from 47\% to 8\% within 20 timesteps (or $\sim$6000 yr) which is the 
frequency of data output used for the simulations. There are no further decreases in f$_{\text{esc,GMC}}$ due to the HII region continually growing in size until it nearly fills the entire 
simulation volume.

\section{Escape Fraction from GMC Populations} \label{Results52}

The results presented above represent the escape fractions from individual GMCs only. Here, we present a simple model to calculate the f$_{\text{esc,tot}}$  --- the 
escape fraction from a population of GMCs that are at different stages in their evolution. We tailor our model to be representative of a dwarf starburst galaxy and also a regular spiral-type galaxy.

Broadly, our model assumes that a new star-forming cloud is birthed every time interval, $\Delta$t. The mass of this cloud is drawn randomly from a GMC mass 
distribution and corresponds to one of the 5 simulations presented above. The GMC is evolved for an assumed lifetime, after which we consider the cloud destroyed. The 
net f$_{\text{esc,tot}}$ and the total SFR of the GMC population at any given time are calculated by summing over the properties of the currently active clouds. The two main 
observational results used by the model are the mass of molecular gas and the depletion time of the galaxy. These are used to determine the appropriate $\Delta$t. We 
note that, over the timescales considered here, the total mass of molecular gas in each model galaxy is assumed to remain constant. This is 
a consequence of how we constructed our model and is described in more detail below.  

We choose a molecular gas mass of 3$\times$10$^9$ M$_{\odot}$, the approximate mass of molecular gas and the cold neutral medium in the Milky Way \citep{Tielens}, for the 
spiral galaxy model and a mass of 10$^8$ M$_{\odot}$ for the starburst dwarf model. This is consistent with local starburst dwarfs studied by \citet{McQuinn}.

We then assume that a new GMC becomes star-forming every time interval $\Delta$t. The mass of this cloud --- which corresponds to one of our 5 models --- is randomly drawn from 
a GMC mass spectrum given by, 

\begin{equation}
dN/dM \propto M^{\alpha},
\end{equation}

\noindent \noindent where $N$ is the number of clouds at a given mass, $M$. We consider two values of $\alpha$. Exponents of 
-1.5 and -2.5 are consistent with the distribution of clouds in the inner MW and M33, respectively \citep{Roso2005}. These values were chosen not only because of observations, but 
because the shallower slope implies most of the total mass in GMCs is contained in high mass clouds and vice versa for the steeper slope. Here, we present the results for the two different mass distributions only for the dwarf model, since the bigger spiral model is much more computationally expensive.

In order to determine the main parameter of our model, $\Delta$t, we use the measured depletion time for our target galaxy types. The depletion time is given by,

\begin{equation}
\tau_{dep} = \frac{M_{\text{H$_{\text{2}}$}}}{\dot{M_*}},
\end{equation}

\noindent \noindent where $M_{\text{H$_{\text{2}}$}}$ is the total mass of molecular gas in the galaxy, and $\dot{M_*}$ is the SFR. Physically, it represents the amount of time, given the current 
mass of H$_{2}$ and the current SFR, it would take to convert all the molecular gas into stars. A gas depletion time for spiral-type galaxies 
is typically $\sim$2.35 Gyr \citep{Bigiel} while $\sim$1 Gyr is representative of dwarf starbursts \citep{McQuinn}. 

\begin{figure*}
\centering
\begin{tabular}{ c c c }
\includegraphics[width=0.45\textwidth]{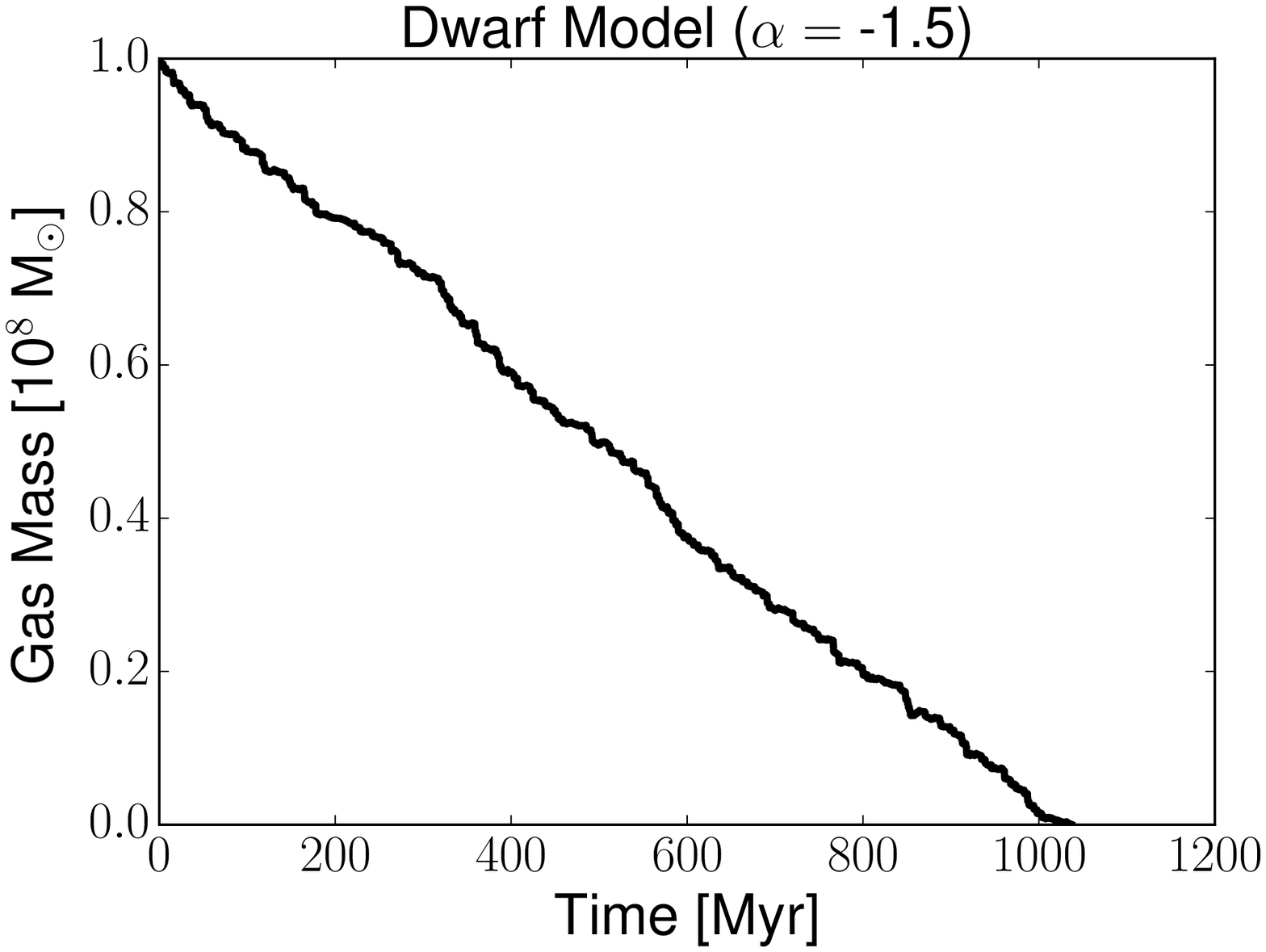} & \includegraphics[width=0.45\textwidth]{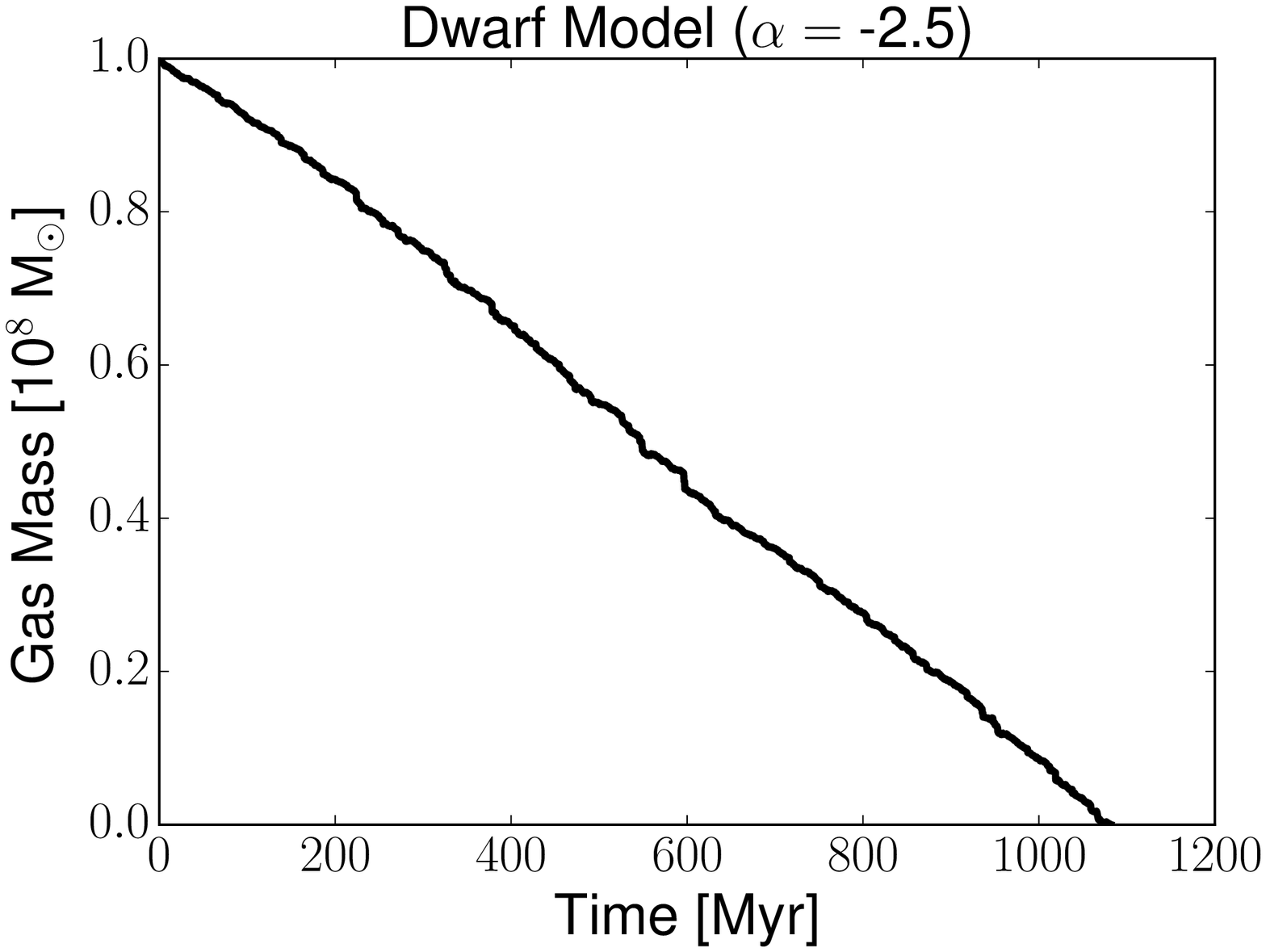} \\
\multicolumn{2}{c}{\includegraphics[width=0.45\textwidth]{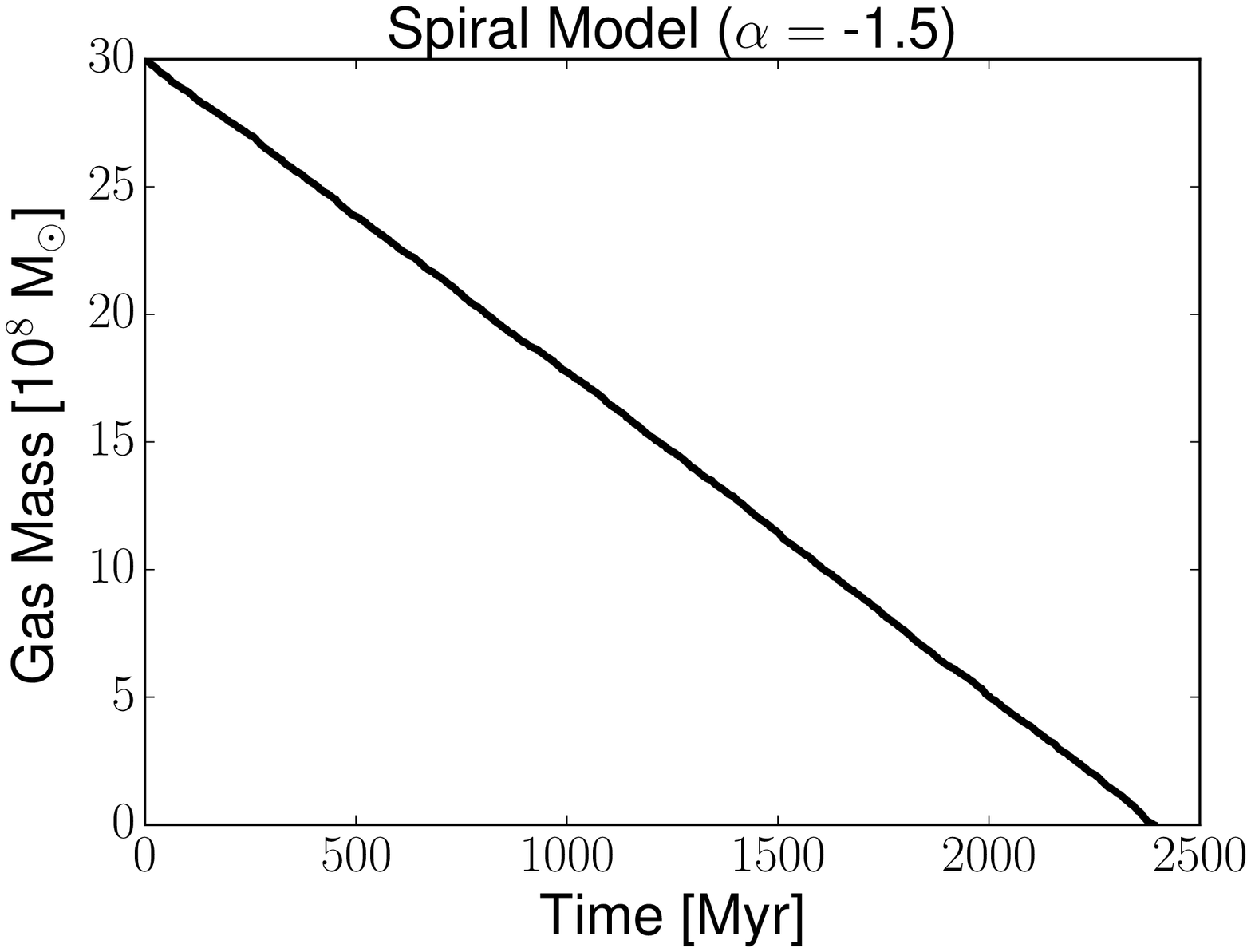} }
\end{tabular}
\caption{The evolution of the mass reservoir out of which GMCs are drawn to determine $\Delta$t for the dwarf model with $\alpha$ $=$ -1.5 (top left), $\alpha$ $=$ -2.5 (top right), and the spiral-type 
model (bottom). We choose the time between GMC formation such that the gas is exhausted after one gas consumption time of $\sim$1 Gyr for dwarf and $\sim$2.35 Gyr for the spiral model.}
\label{fig:reservoir}
\end{figure*}

The appropriate $\Delta$t is found through the following steps:
1) We start with the total amount of molecular gas within the model galaxy, referred to as the 'reservoir'. 2) A new star-forming GMC appears every time interval $\Delta$t. We start with an estimate of this parameter. 
3) We evolve each cloud for an assumed lifetime (see below) after which we consider the cloud destroyed. After the clouds are destroyed, their stars are considered part of the field 
population and are not included in the calculation of f$_{esc,tot}$. 4) Any gas in the cloud that has not been converted to stars is returned to the mass reservoir from which clouds are drawn. 5) We continue forming new clouds until 
the reservoir of molecular gas is exhausted.
 
Through iterative trials, we are able to converge on the $\Delta$t that ensures the initial amount of molecular gas is converted to stars in one depletion time. We note that 
we are not suggesting that all the molecular gas is consumed in a depletion time in a real galaxy, since molecular gas is constantly being formed from atomic gas. Additionally, there is accretion of fresh gas from outside the galaxy. We have adopted the above procedure to be faithful to the definition of the depletion time which is an \textit{instantaneous} measurement of the time it 
would take to convert a galaxy's molecular gas to stars given the current SFR. Our method for determining $\Delta$t ensures we are consistent with this definition. Since $\Delta$t 
remains constant throughout the evolution of our models, we are inherently assuming that the depletion time, and therefore the molecular gas content and the global star formation
properties, of the galaxies do not appreciably change.  

The procedure outlined above results in $\Delta$t values of 0.65 Myr for the dwarf starburst model with $\alpha$ $=$ -1.5, 0.12 Myr for the dwarf starburst model with $\alpha$ $=$ -2.5,
and 52 Kyr for the spiral galaxy model. 

With these values in hand, we complete the final model run by repeating many of the steps already described. Namely, new star-forming clouds are drawn from the adopted GMC 
mass distribution every time interval $\Delta$t and the clouds are evolved for an assumed lifetime after which we consider them destroyed. The length of time we evolve our model 
is a free parameter but overall its effect is minimal as long as it significantly exceeds the lifetime of our individual GMC models. This is because shortly after starting the model, an equilibrium is established between the number of clouds starting 
and stopping star formation resulting in a roughly constant number of GMCs with time. Since an element of stochasticity is introduced via randomly sampling the mass distribution, 
we have chosen to evolve all models through one full depletion time. The figures presented below indicate that this is sufficient for capturing the behaviour of our model.  
   
We take the lifetimes of the 10$^4$, 5$\times$10$^4$ and 10$^5$ M$_{\odot}$ GMCs to be 5 Myr. As shown in \citet{Howard2017-2}, these clouds are nearly entirely disrupted by 
radiative feedback by this time. The more massive clouds, however, show no evidence of large-scale disruption. We therefore adopt a lifetime of 10 Myr for these clouds 
for two reasons. 

Firstly, this value is roughly consistent with the observed lifetimes of GMCs 
in M33 \citep{Corbelli}. The authors, using a sample of 566 GMCs complete down to $\sim$5$\times$10$^4$ M$_{\odot}$, were able to compare the relative frequencies of clouds 
with no evidence of star formation, clouds with evidence of embedded star formation, and clouds hosting HII regions. They found that the time between cloud assembly and when 
the first clusters break through the cloud is 14.2 Myr. Since we do not model the formation stage of the GMCs --- they are considered to be assembled and at the verge of star formation 
when the simulations begin --- 10 Myr is an appropriate lifetime. 

Secondly, we choose a value of 10 Myr for the massive clouds because we make the assumption of a constant f$_{\text{esc,GMC}}$ and SFR after the $\sim$5 Myr of evolution presented in 
Section 3. The final SFR and f$_{\text{esc,GMC}}$ from the simulations are taken for the portion of the cloud evolution past 5 Myr. We adopt a constant SFR because, as shown in \citet{Howard2017}, the 
SFRs of the massive clouds rise to an approximately constant value after $\sim$2.5 Myr and show no evidence for a decreasing SFR at late times. Based on these final SFRs, the 
entirety of the unused gas in the simulation would be consumed at approximately 12 Myr. Therefore, choosing a lifetime of 10 Myr ensures that there is still gas present at the time of 
cloud dispersal which justifies the use of a relatively low f$_{\text{esc,GMC}}$ that extends beyond the end of the simulations. Moreover, a low f$_{\text{esc,GMC}}$ for 
massive clouds at times $>$5 Myr is consistent with the simulations of \citet{Rahner} which include the combined effects of stellar winds and radiative feedback. 

At any given time, we know the total population of clouds, which were randomly drawn from an assumed GMC powerlaw mass distribution, and where they are in their respective evolutionary histories. 
Since the clouds are formed at different times, the population of GMCs will necessarily contain clouds that are nearing the end of their lifetime while others will be just 
starting star formation. The instantaneous number of photons being generated by the clusters, and the fraction of those that escape the cloud, are known for each GMC at all times.
The escape fraction from the ensemble of clouds is then the instantaneous
number of photons escaping all clouds divided by the total number of photons being generated by all clusters. Mathematically, it is expressed as,
\begin{equation}\label{fesc}
f_{\text{esc,tot}} = \frac{\sum_{i} N_{\text{esc},i}}{\sum_{i} N_{\text{tot},i}},
\end{equation}

\noindent \noindent where $N_{\text{esc},i}$ is the number of photons escaping from cloud $i$, and $N_{\text{tot},i}$ is the total number of photons being produced by the stars in cloud $i$. The 
SFR is also calculated by summing the instantaneous SFRs of the individual clouds.

\begin{figure*}
\centering
\begin{tabular}{ c }
\includegraphics[width=0.9\textwidth]{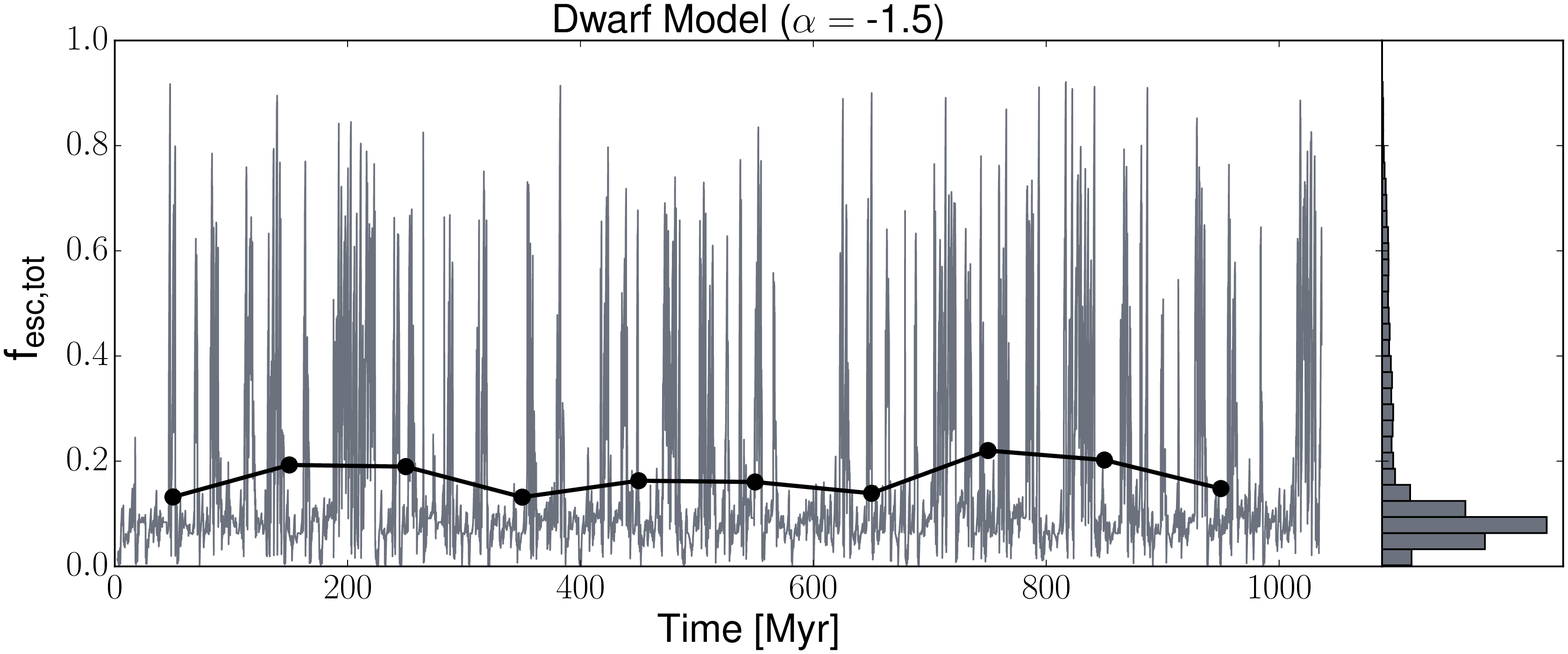} \\
\includegraphics[width=0.9\textwidth]{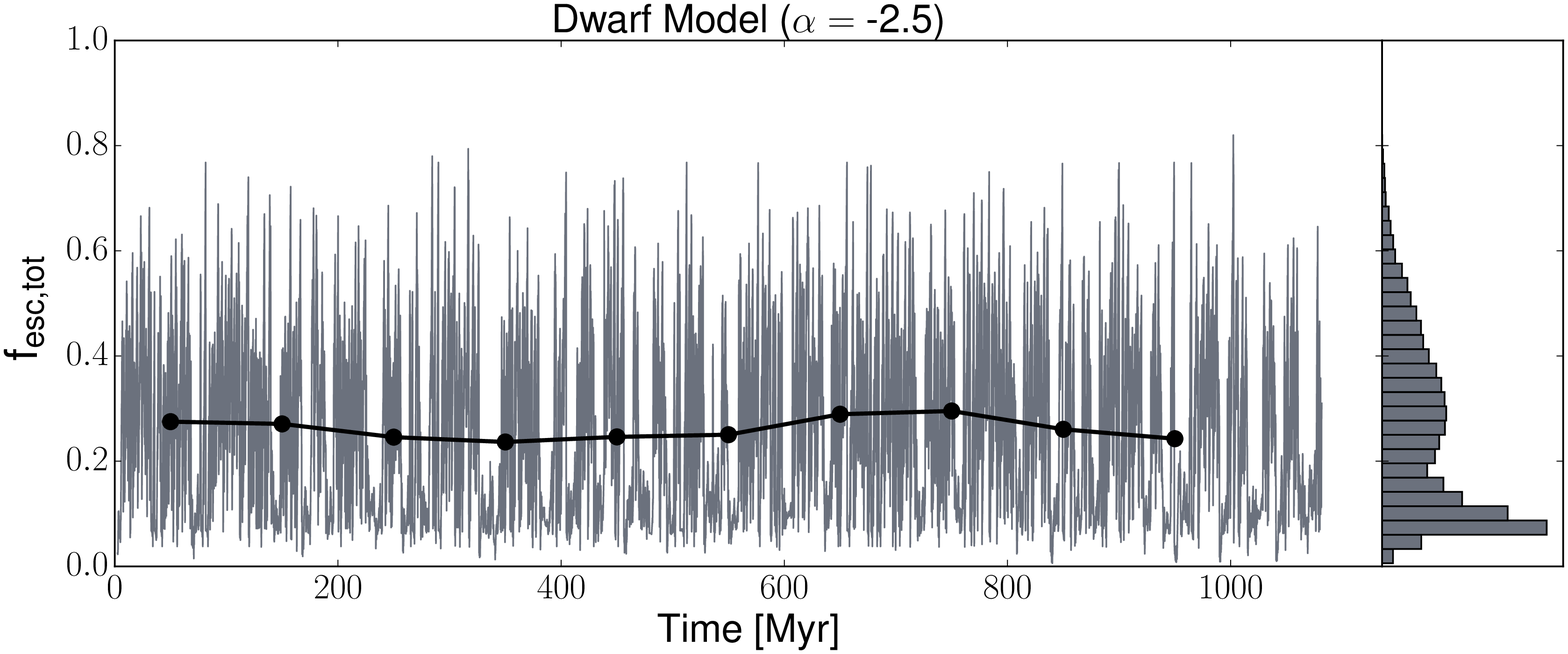} \\
\includegraphics[width=0.9\textwidth]{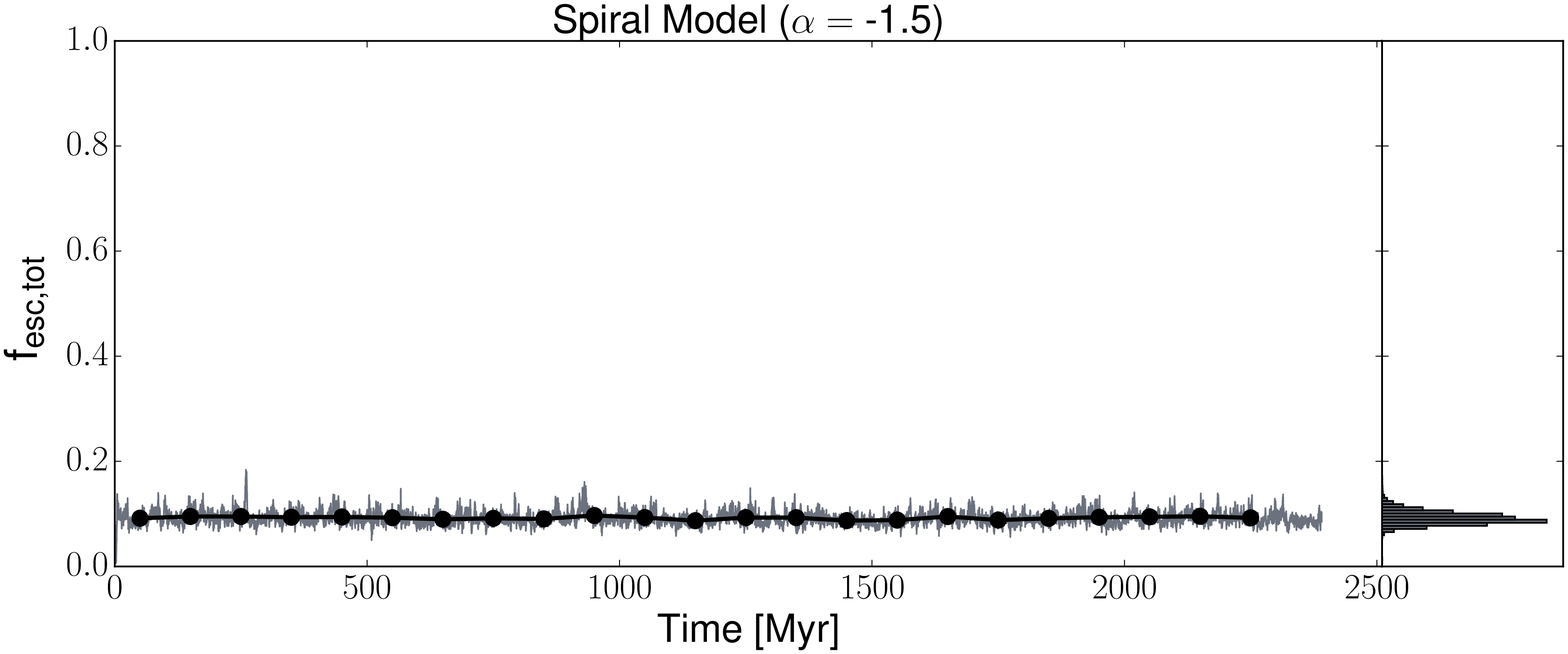}
\end{tabular}
\caption{The escape fraction, f$_{\text{esc,tot}}$, for the two starburst dwarf models (top and middle) and the spiral-type galaxy model (bottom) as it evolves over time. Black circles represent the average 
over 100 Myr timescales. The histogram, plotted with the same vertical scale as the left plot, represents the distribution of f$_{\text{esc,tot}}$ over each galaxy's history.}
\label{fig:escape}
\end{figure*}

We start by showing the evolution of the reservoir gas mass, used to determine the appropriate $\Delta$t values, in Figure \ref{fig:reservoir}. The dwarf starburst models begins with 10$^8$ M$_{\odot}$ which is fully converted to 
GMCs in 1.04 Gyr and 1.08 Gyr for the cases with $\alpha$ $=$ -1.5 and -2.5 respectively. The spiral galaxy model (right) uses 3$\times$10$^9$ M$_{\odot}$ in 2.39 Gyr. 

The resulting f$_{\text{esc,tot}}$ for the two dwarf models (top and middle) and the spiral model (bottom) are shown as a function of time in Figure \ref{fig:escape}. The averages over a 100 Myr 
timescale are shown by the solid black points. The histogram represents the normalized distribution of f$_{\text{esc,tot}}$ over the entire history of each model. The vertical scale
is identical for the time evolution plot and the histogram. 

We see that the dwarf model with $\alpha$ $=$ -1.5 is characterized by a highly variable f$_{\text{esc,tot}}$ that ranges from 0-90\%. While there are times of high f$_{\text{esc,tot}}$, the typical 
escape fraction, as shown by the histogram to the right of the plot, is small. The peak of the f$_{\text{esc,tot}}$ histogram occurs at 7.8\%.

The middle panel of Figure \ref{fig:escape} demonstrates that varying the slope of the GMC mass distribution can alter the time evolution of f$_{\text{esc,tot}}$. While the most 
likely value of f$_{\text{esc,tot}}$ remains roughly constant at 7.4\% when considering $\alpha$ $=$ -2.5, the probability of having a higher f$_{\text{esc,tot}}$ at any given time is increased, as 
demonstrated by the secondary peak that occurs at f$_{\text{esc,tot}}$ $\sim$ 30\%. This is attributed to the different population of GMCs in each case. For the shallower slope ($\alpha$ $=$ -1.5), 
more mass is contained in the most massive clouds which have a correspondingly low f$_{\text{esc,GMC}}$ (see Figure \ref{fig:grid}). Due to the large stellar population in the massive 
clouds, they contribute strongly to the denominator in Equation \ref{fesc}. However, only a small fraction of the UV photons escape the host GMC. This effectively weights f$_{\text{esc,tot}}$
to smaller values. On the other hand, the case with $\alpha$ $=$ -2.5 will have a more mass contained 
in these low to intermediate mass clouds and therefore a higher f$_{\text{esc,tot}}$. Moreover, since $\Delta$t is smaller for $\alpha$ $=$ -2.5 --- a higher fraction of low mass 
GMCs means that clouds need to form more often in order to exhaust the gas reservoir in the same amount of time --- there are more GMCs present at any given time which increases 
the likelihood of having 10$^{4-5}$ M$_{\odot}$ clouds near the end of their life cycle at times when f$_{\text{esc,GMC}}$ is high.

\begin{figure*}
\centering
\begin{tabular}{ c c }
  \includegraphics[width=0.5\textwidth]{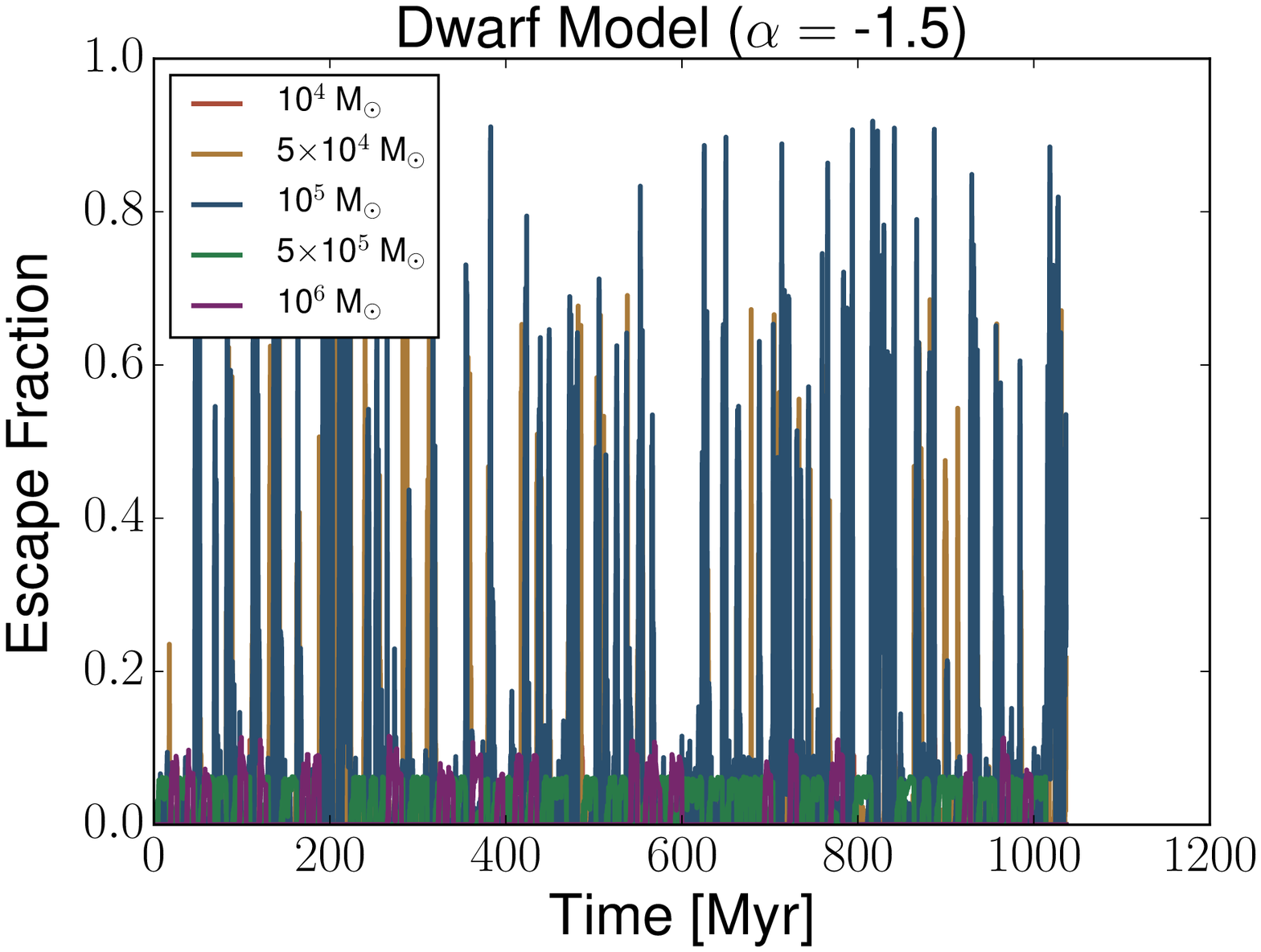} &  \includegraphics[width=0.5\textwidth]{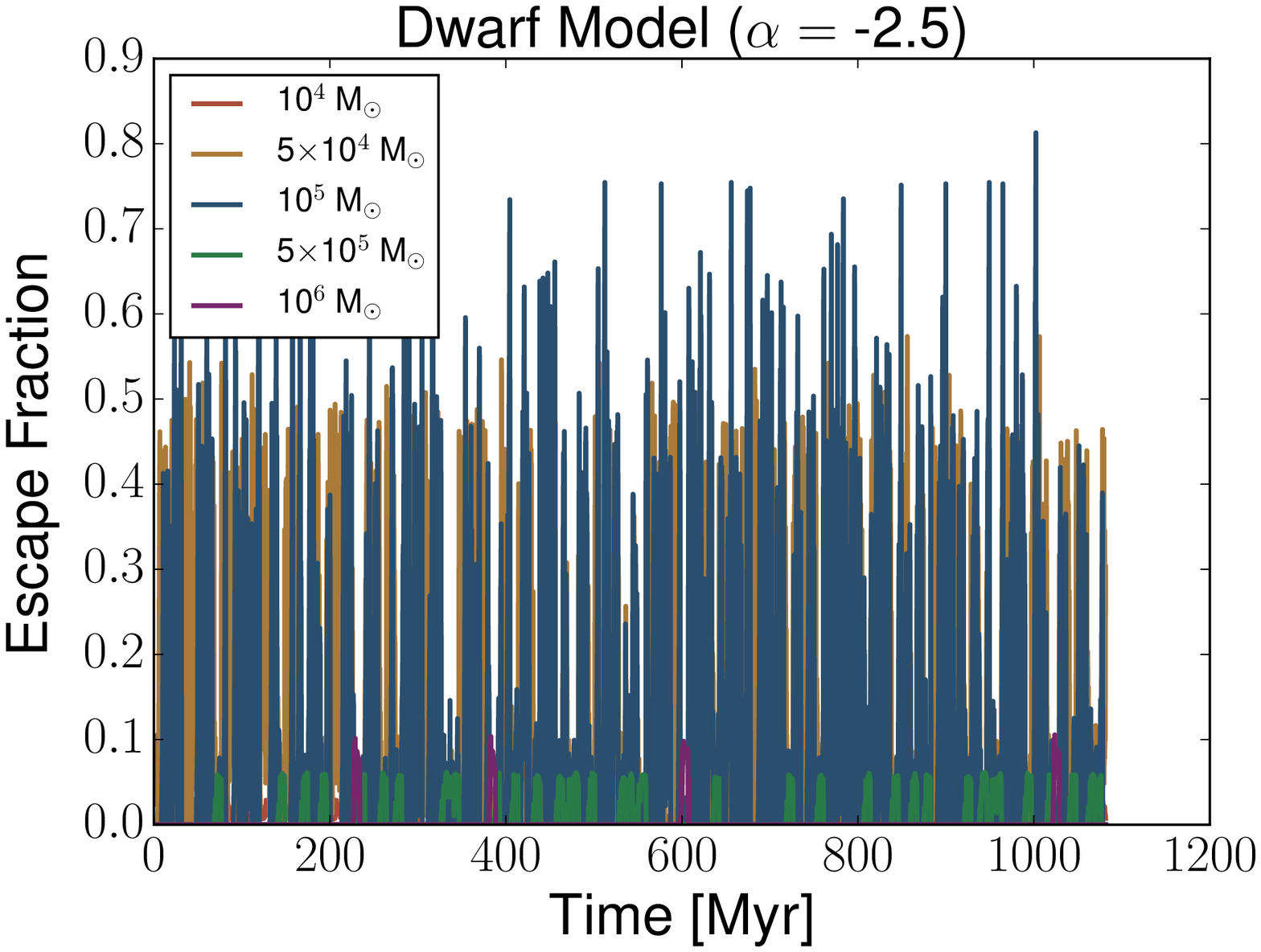}
\end{tabular}
\caption{The contributions to f$_{\text{esc,tot}}$ by each of the 5 cloud masses. Note that there can be many GMCs with the same mass at any given time. The sum of these 
curves represents the total f$_{\text{esc,tot}}$ presented in Figure \ref{fig:escape}.}
\label{fig:contributions}
\end{figure*}

Our claim that the intermediate mass clouds with high f$_{\text{esc,tot}}$ are responsible for the peaks seen in Figure \ref{fig:escape} is supported by Figure \ref{fig:contributions}. Here, 
we separate f$_{\text{esc,tot}}$ into the contributions from the different GMC masses. The sum of these curves returns the f$_{\text{esc,tot}}$ discussed above. It is clear 
from the Figure \ref{fig:contributions} that the 5$\times$10$^4$ and 10$^5$ M$_{\odot}$ GMCs are responsible for the peaks of f$_{\text{esc,tot}}$. The more massive GMCs, on the other hand, only 
contribute up to 10\% of total photons that are escaping from the population of molecular clouds despite hosting a larger stellar population. The low f$_{\text{esc,GMC}}$ from these 
clouds (see Figure \ref{fig:grid}), combined with their low frequency compared to less massive clouds, limits their contribution to the total f$_{\text{esc,tot}}$. 

The lower panel of Figure \ref{fig:escape} shows f$_{\text{esc,tot}}$ for the spiral model. Compared to the dwarf models, there is significantly less variation in f$_{\text{esc,tot}}$ which 
only ranges from $\sim$5-18\%. The most common f$_{\text{esc,tot}}$ (i.e. the peak of the histogram) is 8.6\%. This is consistent with f$_{\text{esc,tot}}$ from both dwarf models, indicating
that an escape fraction of $\sim$8\% is a robust result regardless of galaxy type or GMC mass distribution. However, the larger variations in the dwarf models means an increased
probability of observing a high f$_{\text{esc,tot}}$ at any given time.

The higher variation in the dwarf model is due to the larger $\Delta$t parameter (i.e. forming new GMCs less frequently). For the dwarf model with $\alpha$ $=$ -1.5, we adopted a $\Delta$t value of 0.65 Myr so there will be only $\sim$8 star-forming GMCs at any given 
time. Due to the stochastic nature of our model, this will inevitably lead to times when several of the GMCs will be either 5$\times$10$^4$ or 
10$^5$ M$_{\odot}$ which have higher f$_{\text{esc,GMC}}$ relative to the other clouds. This results in the peak f$_{\text{esc,tot}}$ of $\sim$90\%. At other times, there will be high 
mass clouds present (5$\times$10$^5$ and 10$^6$ M$_{\odot}$) that generate a large number of photons but have a low corresponding f$_{\text{esc,GMC}}$. This skews f$_{\text{esc,tot}}$ to small values.

In contrast, the shorter $\Delta$t value of 52 kyr for the spiral model means there will be at least $\sim$100 clouds present at a given time. This limits the impact of stochastic sampling 
and results in a more time-insensitive distribution of GMC masses. The significantly higher number of GMCs present at any time means the high f$_{\text{esc,GMC}}$ from 5$\times$10$^4$ and 10$^5$ M$_{\odot}$ are less pronounced. 
Nonetheless, a similar analysis to Figure \ref{fig:contributions}, which is not shown here due to its similarity to the Figures already presented, indicates that 
these two clouds masses are still the dominant contributors to the total f$_{\text{esc,tot}}$ for the spiral model.   

It is worth noting that the results presented in Figure \ref{fig:escape} share qualitative properties with galactic f$_{\text{esc}}$ obtained via cosmological zoom-in 
radiation-hydrodynamic simulations. The 
galactic f$_{\text{esc}}$ values are also found to vary by several orders of magnitude over timescales of $\sim$10 Myr \citep{KimmCen,Trebitsch}. This is related to the timescale for massive stars 
to be born and die, suggesting that Supernovae play a controlling role in galactic escape fractions. For mini-haloes in the early Universe, however, heating 
due to photoionization has been shown to be the main process responsible for the efficient escape of ionizing photons \citep{Kimm2017}. In our case, we obtain a similar timescale from the 
assumed lifetimes of our GMCs. Overall, the turbulent nature of GMCs, and the ISM as a whole, plays a cental role in controlling the escape of UV photons.

\subsection{Comparison to M33}

We note that the number of star-forming clouds in the spiral model is less than the number found in M33 by \citet{Corbelli}. In that work, the authors identify a total of 
566 GMCs in M33 which represents a complete sample down to $\sim$5$\times$10$^4$ M$_{\odot}$. A large fraction (32\%) of the GMCs, however, show no evidence for star formation. 
A total of 369 GMCs show either embedded star formation or contain visible HII regions. Our model, therefore, has approximately 4 times fewer star-forming clouds than 
their sample. This is attributed to the high SFEs of our individual GMC simulations which range from 16\% to 21\%. As discussed in \citep{Howard2017}, this is likely due to 
not including other forms of feedback during the early phases of cluster formation (i.e. stellar winds). By not including these additional forms of early stellar feedback, the escape fractions presented above should be interpreted as lower limits. Since our SFEs are higher than those observed in local GMCs, we 
require fewer clouds overall to fully convert the molecular gas to stars in one depletion time.  

\subsection{Comparison with Observations of Galactic Star Formation Rates}

\begin{figure*}
\centering
\begin{tabular}{ c }
  \includegraphics[width=0.9\textwidth]{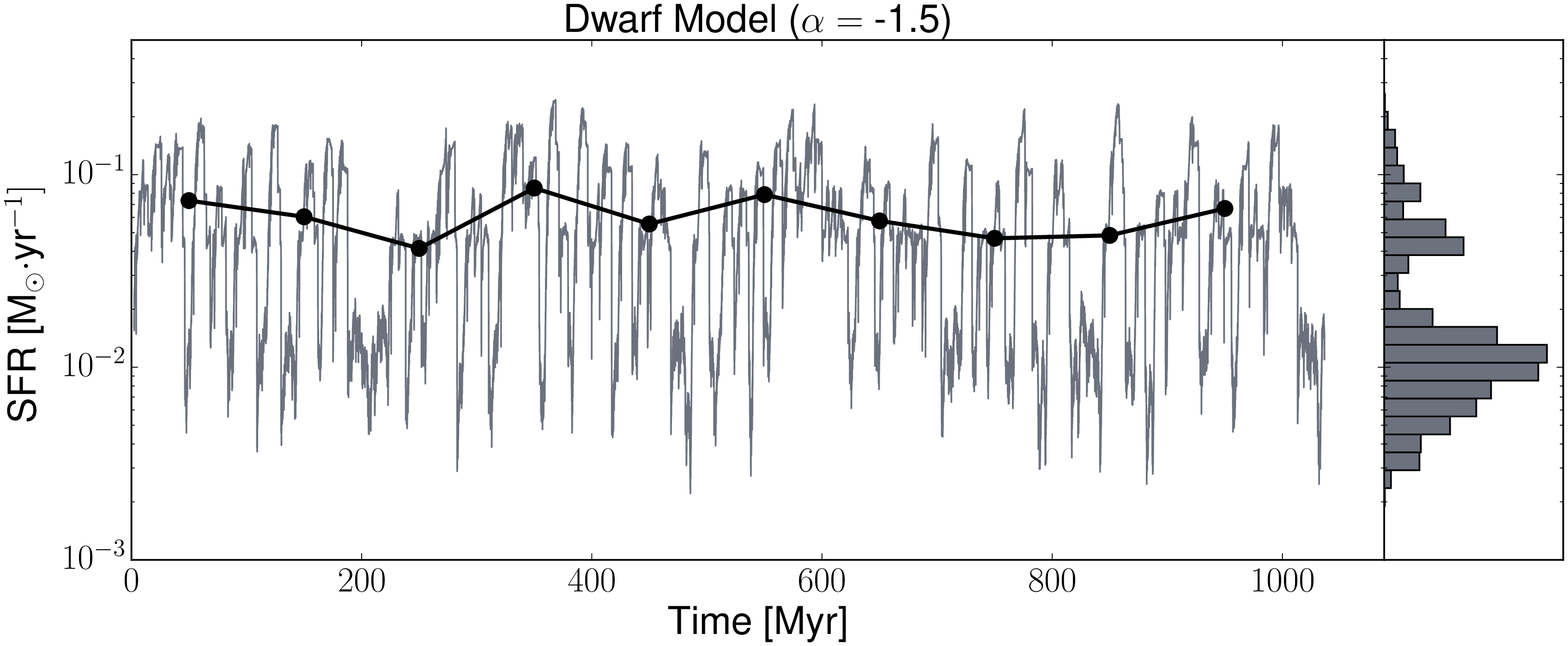} \\
  \includegraphics[width=0.9\textwidth]{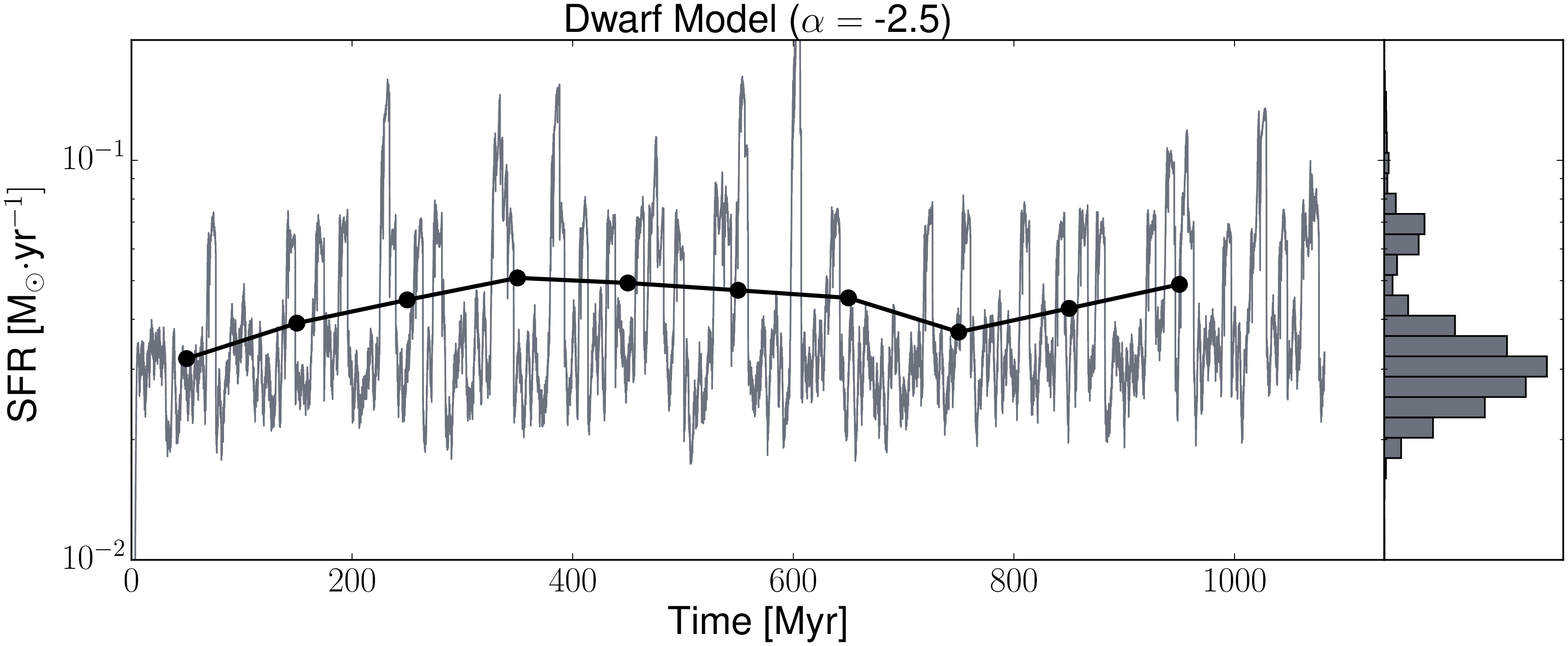} \\
  \includegraphics[width=0.9\textwidth]{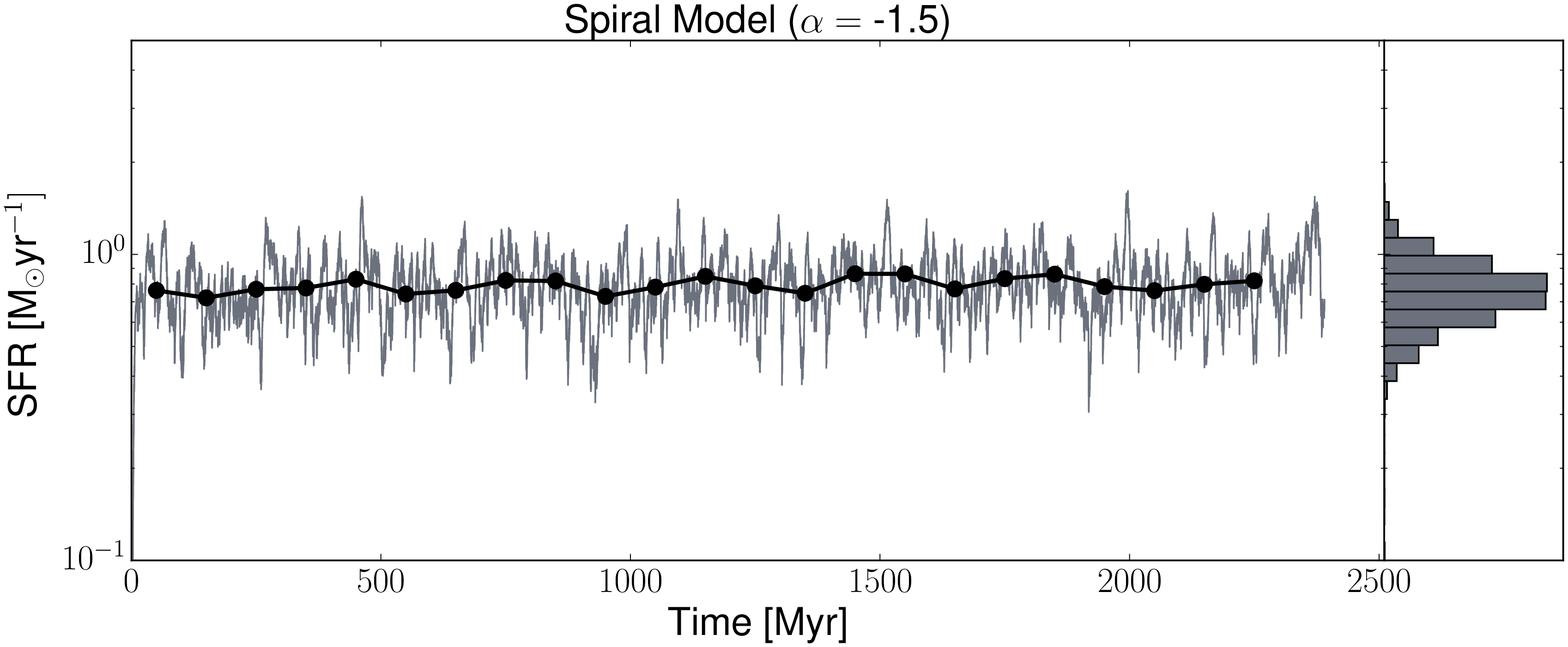}
\end{tabular}
\caption{The evolution of the star formation rate (SFR) for the dwarf starburst models and the spiral-type galaxy model. As in Figure \ref{fig:escape}, the black dots represent the average
values in 100 Myr bins and the histogram shows the overall distribution of SFRs over each model's history.}
\label{fig:SFRs}
\end{figure*}

To compare with observations of starburst dwarf and spiral-type galaxies, we plot the SFRs, given by the summed instantaneous SFRs of the cloud population, in Figure \ref{fig:SFRs}.

The SFR for the dwarf model with $\alpha$ $=$ -1.5 ranges between 2$\times$10$^{-3}$ and 0.3 M$_{\odot}$yr$^{-1}$. The most likely SFR from the histogram is 1.2$\times$10$^{-2}$ M$_{\odot}$yr$^{-1}$. 
Changing the GMC mass distribution slope to $\alpha$ $=$ -2.5 results in a similar range (10$^{-2}$ to 0.2 M$_{\odot}$yr$^{-1}$) and a slightly larger peak value of 3.1$\times$10$^{-2}$ M$_{\odot}$yr$^{-1}$.
We can directly compare to observed SFRs, and inferred star formation histories (SFHs), of starburst dwarfs from \citet{McQuinn} who were able to reconstruct the 
SFHs of local dwarfs using a combination of stellar photometry and stellar evolution models. The results from this analysis 
indicate peak SFRs which range from 5.2$\times$10$^{-4}$ to 9.7$\times$10$^{-1}$ M$_{\odot}$yr$^{-1}$ depending on the galaxy studied. Their average peak SFR was 0.13 M$_{\odot}$yr$^{-1}$. The results presented here are consistent with several of their observed galaxies but we note that the star formation histories of dwarf starbursts vary significantly. Our results 
are also consistent with the observations of \cite{Weisz} who measured the average SFRs in dwarfs of various morphologies over their entire history. A similarly large spread in 
SFRs between galaxies is also found. 

The temporal variation of the SFRs for the dwarf model is also consistent with the inferred SFHs. The results presented in \citet{McQuinn} and \citet{McQuinn2} demonstrate 
variations of up to an order of magnitude in the SFRs of dwarf starbursts over timescales of 10-20 Myr. The combination of our GMC scale physics and the stochastic sampling 
of a GMC mass distribution naturally reproduce this feature.

Similar to f$_{\text{esc,tot}}$ presented in Figure \ref{fig:escape}, the spiral model shows less variation in the SFR compared to the dwarf model. The SFR varies from $\sim$0.3 to 1.5 
M$_{\odot}$yr$^{-1}$. The peak of the SFR histogram occurs at 0.73 M$_{\odot}$yr$^{-1}$.

The SFR from our model is within the range of local spiral galaxies. The SFR of the Milky Way, which was used as a basis for 
choosing the initial reservoir mass for the spiral model, is $\sim$1.65 M$_{\odot}$yr$^{-1}$ \citep{MilkyWay} --- approximately a factor of 2 higher than the peak SFR in Figure \ref{fig:SFRs}. The SFRs measured for local, normal spiral galaxies are typically between 
0.5-10 M$_{\odot}$yr$^{-1}$ \citep{Gao2004}. Our results fall well within this observed range.

Therefore, by utilizing only a suite of 5 GMC mass models under the assumption of a constant GMC formation rate, we have been able to reproduce the 
behaviour of the SFRs in dwarfs and spirals. This is likely due to using the gas consumption time as a main observational constraint in our model since it inherently depends on the SFR. 

\subsection{f$_{\text{esc,tot}}$ - SFR relation}

\begin{figure*}
\centering
\begin{tabular}{ c c }
\includegraphics[width=.45\textwidth]{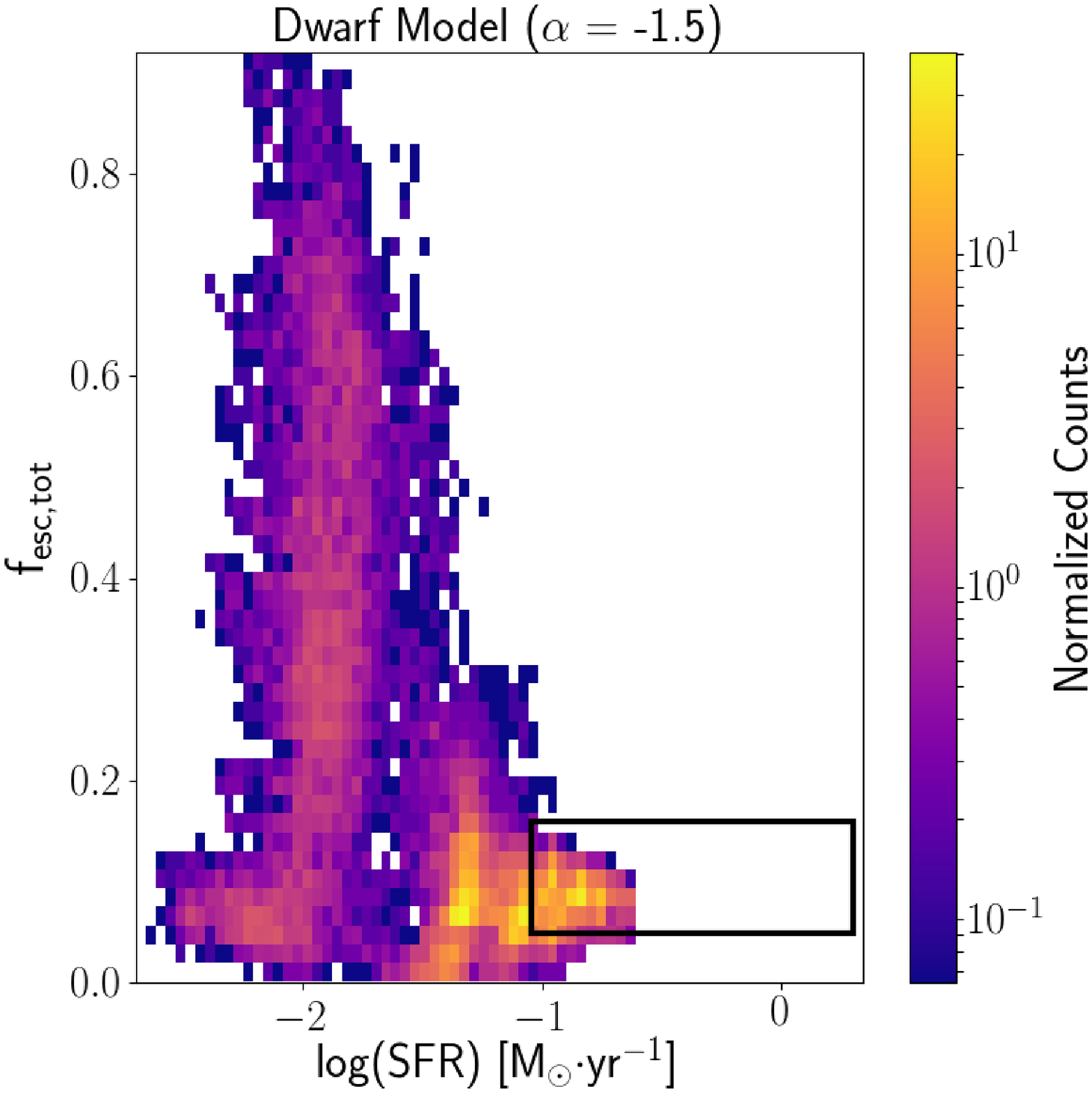} & \includegraphics[width=.45\textwidth]{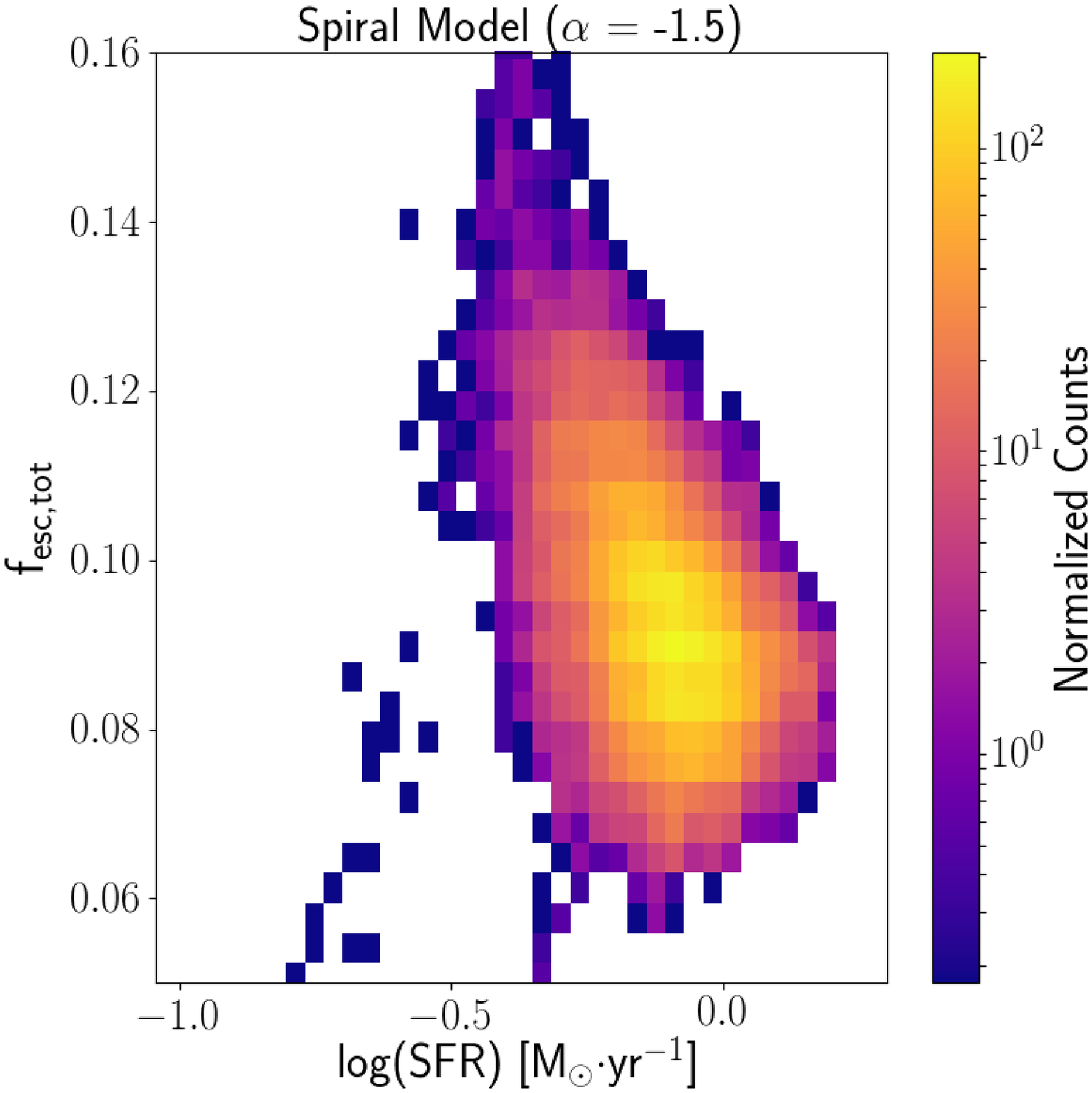} \\
\includegraphics[width=.45\textwidth]{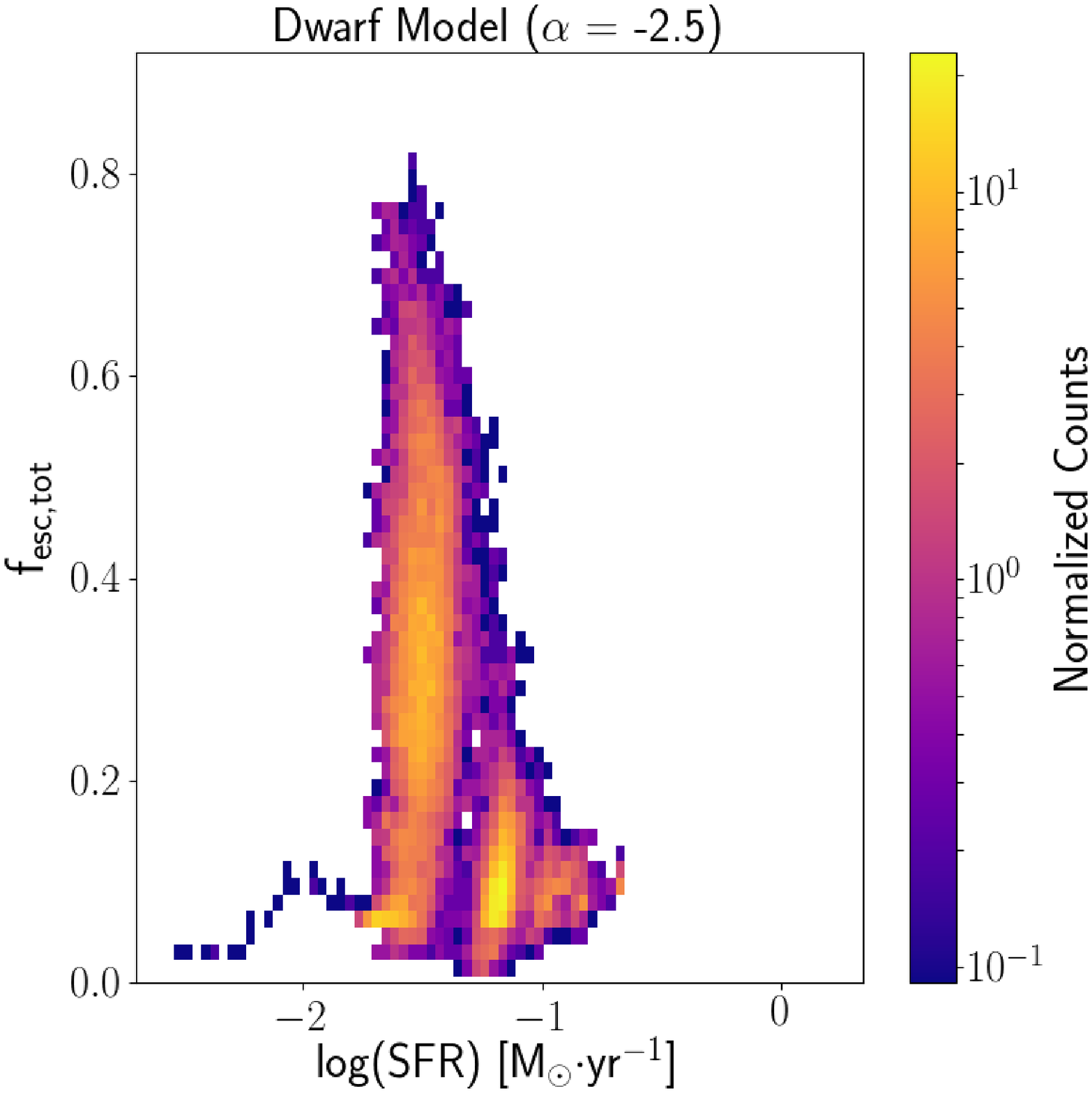}
\end{tabular}
\caption{2D histogram showing the correlation between f$_{\text{esc,tot}}$ and SFR for the two dwarf models (top left and bottom) and the spiral model (top right). The black box in the dwarf model with $\alpha$ $=$ -1.5 represents the range of the spiral model plot. 
The colours represent the normalized counts in each bin.}
\label{fescSFR}
\end{figure*}

We conclude by combining the results from the two previous Sections in order to examine the relation between f$_{\text{esc,tot}}$ and SFR. This may allow for f$_{\text{esc,tot}}$, an observationally difficult quantity to constrain, to be obtained from measurements of the SFR. We have plotted 2D histograms in Figure \ref{fescSFR} for 
the two dwarf models (top left and bottom) and the spiral model (top right). For clarity, a black box is drawn on the dwarf model with $\alpha$ $=$ -1.5 which represents the range of the spiral model plot. The colours represent 
the normalized number of counts in each bin.

The dwarf models show two distinct regions --- a grouping at low SFR that covers a large range of f$_{\text{esc,tot}}$, and a grouping at higher SFR but low ($\sim$10\%) f$_{\text{esc,tot}}$. Overall, this indicates that higher SFRs are typically associated with lower escape 
fractions. The region at high SFR and low f$_{\text{esc,tot}}$ corresponds to times when massive 
(5$\times$10$^5$ and 10$^6$ M$_{\odot}$) GMCs are actively star-forming. These GMCs have the highest SFRs and therefore significantly increase the total SFR of the population of 
clouds. However, as discussed above, f$_{\text{esc,GMC}}$ is low in the massive clouds. Since these clouds generate a large number of UV photons, of which only a small fraction 
escapes the host GMC, this weights f$_{\text{esc,tot}}$ to small values. The grouping at low SFR. on the other hand, corresponds to times when only low to intermediate mass 
clouds are present. These GMCs have lower SFRs and f$_{\text{esc,GMC}}$ that approach 100\%. This results in the increased range of f$_{\text{esc,tot}}$.

The claim that the distinct regions in this diagram are caused by the presence, or lack of, massive clouds is supported by comparing the plots of the two dwarf models. The model with an $\alpha$ 
of -1.5 shows a stronger concentration towards the bottom right of the plot. The model with $\alpha$ $=$ -2.5, on the other hand, has a more balanced distribution between 
the two regions. Since the model with the shallower slope (-1.5) has a relatively larger fraction of massive clouds, a stronger concentration at high SFR and low 
f$_{\text{esc,tot}}$ is expected. Massive GMCs are more infrequently formed in the case with $\alpha$ $=$ -2.5 so the low SFR and high f$_{\text{esc,GMC}}$ region of the plot is 
more populated.

The spiral model shows a more continuous distribution of data points. Comparing the axis limits to the dwarf models (the black box indicates the range of the spiral model results),
we see that all points are in the high SFR and low f$_{\text{esc,tot}}$ region. This is because the spiral model has significantly more clouds at any given time compared to 
the dwarf models, thereby increasing the likelihood of having massive GMCs present. Like the dwarf models, a slight negative trend with large scatter is 
present. 

Overall, the above results indicate that, within a given class, galaxies with higher than average SFRs should statistically have lower escape fractions from their 
population of GMCs. When measurements of f$_{\text{esc,tot}}$ are performed on large numbers of galaxies, this prediction can be directly tested. 

\section{Discussion and Conclusions}

The escape of UV photons from GMCs which host massive stars drives many crucial astrophysical processes, from determining the chemical, thermal, and ionization state of the ISM 
by contributing to the ISRF \citep{Draine2011} at present days to participating in cosmic reionization \citep{Wise2014} at high redshift. Estimates of the escape fraction from an entire galaxy vary 
significantly --- the distribution of dense gas being a main factor in controlling the escape fraction.

In order to place constraints on the escape fraction from both individual and populations of molecular clouds, we present the UV photon escape fraction from a suite of GMC models simulated 
using the FLASH code. The simulations, taken from \citep{Howard2017-2}, consisted of 5 GMCs with masses of 10$^4$, 5$\times$10$^{4}$, 10$^5$, 5$\times$10$^{6}$, and 
10$^6$ M$_{\odot}$. All clouds had the same initial density of 100 cm$^{-3}$ and an initial virial parameter of 3. Sink particles, coupled with a custom subgrid model, are used 
to model cluster formation and radiative transfer is included via a raytracing scheme.

We note that the escape fractions presented in this work should be interpreted as lower limits. We do not include the effects of stellar winds or supernovae in our simulations which have been 
shown to significantly alter the density structure surrounding massive stars \citep{Dale2008, Rahner}. The momentum imparted by stellar winds can remove gas from a cluster's 
surroundings, resulting in low density regions which are easily ionized thereby allowing more UV photons to escape the cloud.

To represent f$_{\text{esc,tot}}$ --- the total escape fraction from a population of GMCs --- we develop a model which forms clouds by randomly sampling a GMC mass distribution over fixed time intervals. Two 
realizations of the model were completed to represent a dwarf starburst galaxy and a normal spiral-type galaxy.

The main input into this model are total molecular gas mass of the galaxy, and the time between cloud formation ($\Delta$t). Masses of 10$^8$ and 
3$\times$10$^{9}$ M$_{\odot}$ are used for the dwarf and spiral models, respectively. A simulated cloud is drawn randomly from a GMC mass distribution 
of $dN/dM\propto$ M$^{\alpha}$ every time interval $\Delta$t, and the cloud is evolved for either 5 Myr for 10$^{4-5}$ M$_{\odot}$ or 10 Myr
for more massive objects. The appropriate values for $\Delta$t are determined by ensuring that the mass of molecular gas 
we have adopted is converted to stars in one depletion time --- taken to be 1 Gyr for dwarfs \citep{McQuinn} and 2.35 Gyr for spirals \citep{Bigiel}. This process is repeated and 
the net f$_{\text{esc,tot}}$ and SFR from the population of GMCs is calculated. For the spiral model, we fix $\alpha$ at -1.5 but complete two realizations --- $\alpha$ $=$ -1.5 and 
-2.5 --- of the dwarf starburst model.

We stress that f$_{\text{esc,tot}}$ represents the escape fraction from a population of GMCs and not from a whole galaxy into the IGM. To fully constrain the latter problem, 
a full treatment of the ISM in galaxies is required including the more diffuse components of the gas, as well as the contribution from field stars. Nevertheless, the results from this work are valuable inputs for 
galactic scale modeling due to the increased resolution and the inclusion of physical processes (eg. radiative feedback) that are typically neglected in larger scale models.

The main conclusions of this work are summarized as follows:

\begin{itemize}

\item Escape fractions from individual clouds (f$_{\text{esc,GMC}}$) vary strongly with time. The final f$_{\text{esc,GMC}}$ values for individual clouds, in order of ascending GMC mass, are 31\%, 90\%, 100\%, 6\%, and 9\%. The high final f$_{\text{esc,GMC}}$ for the 5$\times$10$^4$ and 
10$^5$ M$_{\odot}$ clouds is due to nearly complete ionization of the GMC at 5 Myr. All models show large fluctuations (up to a factor of 6) over small timescales. 
These fluctuations are attributed to dynamic HII regions within the highly filamentary molecular clouds which grow and shrink rapidly depending on the local conditions surrounding the clusters.

\item The escape fraction from our dwarf starburst models fluctuate from values near zero to 90\%. The typical value for f$_{\text{esc,tot}}$ is 7-8\% for both values of $\alpha$, but the 
model with $\alpha$ $=$ -2.5 shows a higher degree of variation. The 5$\times$10$^4$ and 10$^5$ M$_{\odot}$ GMCs, which have the highest f$_{\text{esc,GMC}}$ at late times, are responsible for peaks 
in f$_{\text{esc,tot}}$. The corresponding SFRs are in the range of 10$^{-3}$ and 0.3 M$_{\odot}$yr$^{-1}$ with typical values of 1.2$\times$10$^{-2}$ and 3.1$\times$10$^{-2}$ M$_{\odot}$yr$^{-1}$ for 
$\alpha$ of -1.5 and -2.5, respectively. These values are consistent with the reconstructed star formation histories 
of starburst dwarf galaxies. The observations of dwarfs also show rapid variations in the SFR over 10-20 Myr timescales which is recovered by our model.

\item The spiral model shows significantly less variation in f$_{\text{esc,tot}}$ (5-18\%) compared to the dwarf models. However, the most likely value of f$_{\text{esc,tot}}$ remains constant (8.6\%)
indicating that, to first order, f$_{\text{esc,tot}}$ can be taken to be 8\% regardless of galaxy type. We find SFRs of $\sim$0.73 M$_{\odot}$yr$^{-1}$
with the highest values reaching 1.5 M$_{\odot}$yr$^{-1}$. This is comparable to the MW \cite[1.65 M$_{\odot}$yr$^{-1}$, see eg.][]{MilkyWay} and is consistent with nearby, spiral-type galaxies which typically 
have SFRs in the range of 0.5 to 10 M$_{\odot}$yr$^{-1}$ \citep{Gao2004}.

\item We find that, for all models, times characterized by a high SFR are typically associated with low ($\sim$10\%) f$_{\text{esc,tot}}$. This is due to the presence of massive, 
star-forming GMCs. At low SFRs, f$_{\text{esc,tot}}$ covers a larger range, reaching up to 90\% for the dwarf models. \\ \\

The success of our model in reproducing the properties of our target objects, in combination with our detailed treatment of GMC scale cluster formation, 
means our f$_{\text{esc,tot}}$ results can provide important constraints for galactic scale simulations. The variations in the escape fractions from one GMC, and from a population as 
a whole, also highlight the importance of a fully self-consistent and highly resolved treatment of star formation in simulations which study the escape fraction of UV photons 
from a galaxy. We will examine the implications of these UV escape fractions from GMCs in galaxies, to the question of cosmic reionization, in a future paper.    

\end{itemize}

\section*{Acknowledgments}

We thank the anonymous referee for their insights and suggestions. We would also like to thank Christine Wilson and James Wadsley for their useful discussion. CSH acknowledges financial support provided by the Natural Sciences and Engineering Research Council (NSERC)
through a Postgraduate scholarship. REP and WEH are supported by Discovery Grants from the Natural Sciences and Engineering Research Council (NSERC) of Canada. RSK acknowledges support from the Deutsche Forschungsgemeinschaft in the Collaborative Research Centre SFB 881 "The Milky Way System" (subprojects B1, B2, and B8) and in the Priority Program SPP 1573 "Physics of the Interstellar Medium" (grant numbers KL 1358/18.1, KL 1358/19.2). RSK furthermore thanks the European Research Council for funding in the ERC Advanced Grant STARLIGHT (project number 339177). The FLASH code was in part developed by the Department of Energy (DOE) supported Alliances Centre for Astrophysical Thermonuclear Flashes
(ASCI) at the University of Chicago. Computations were performed on the gpc supercomputer at the SciNet HPC
Consortium. SciNet is funded by: the Canada Foundation for Innovation under the auspices of Compute Canada; the
Government of Ontario; Ontario Research Fund - Research Excellence; and the University of Toronto.

\bibliography{howard_paper4}
\bsp

\label{lastpage}

\end{document}